\documentclass[fleqn,usenatbib]{mnras}

\footnotesize
\newdimen\minuswidth    %define @ width of minus sign for tables
\setbox0=\hbox{$-$}
\minuswidth=\wd0
\catcode`@=\active
\def@{\kern\minuswidth}
\newdimen\digitwidth    %define ! a one digit width for tables
\setbox0=\hbox{\rm0}
\digitwidth=\wd0
\catcode`!=\active
\def!{\kern\digitwidth}
\normalsize
\def\lesssim{\mathrel{\hbox{\rlap{\hbox{\lower4pt\hbox{$\sim$}}}\hbox{$<$}}}}
\def\gtrsim{\mathrel{\hbox{\rlap{\hbox{\lower4pt\hbox{$\sim$}}}\hbox{$>$}}}}

\usepackage{newtxtext,newtxmath}
\usepackage[T1]{fontenc}
\usepackage{ae,aecompl}

\usepackage{graphicx}	% Including figure files
\usepackage{amsmath}	% Advanced maths commands
\usepackage{amssymb}	% Extra maths symbols
\usepackage{pdflscape}
\usepackage{cleveref}
\usepackage{hyperref}

\title[Long-term timing of the pulsars in 47 Tucanae]{Long-term observations of the pulsars in 47 Tucanae - II. Proper motions, accelerations and jerks}

\author[Paulo C. C. Freire et al.]{\parbox{\textwidth}{
P. C. C. Freire$^{1}$\thanks{E-mail: pfreire at mpifr-bonn.mpg.de},
A. Ridolfi$^{1}$,
M. Kramer$^{1}$,
C. Jordan$^{2}$,
R. N. Manchester$^{3}$,
P. Torne$^{1,4}$,
J. Sarkissian$^{3}$,
C. O. Heinke$^{5}$,
N. D'Amico$^{6,7}$,
F. Camilo$^{8}$,
D. R. Lorimer$^{9, 10}$ and
A. G. Lyne$^{2}$
}
\vspace{0.4cm}\\
\parbox{\textwidth}{
$^{1}$Max-Planck-Institut f\"ur Radioastronomie, Auf dem H\"ugel 69, D-53121 Bonn, Germany\\
$^{2}$Jodrell Bank Centre for Astrophysics, School of Physics and Astronomy, The University of Manchester, Manchester M13 9PL, UK\\
$^{3}$CSIRO Astronomy and Space Science, Australia Telescope National Facility, Box 76, Epping, NSW 1710, Australia\\
$^{4}$Instituto de Radioastronom\'ia Milim\'etrica, Avda. Divina Pastora 7, N\'ucleo Central, 18012, Granada, Spain\\
$^{5}$Physics Department, University of Alberta, 4-183 CCIS, Edmonton, AB T6G 2G7, Canada\\
$^{6}$Osservatorio Astronomico di Cagliari, INAF, via della Scienza 5, I-09047 Selargius (CA), Italy\\
$^{7}$Dipartimento di Fisica, Universit\`a degli Studi di Cagliari, SP Monserrato-Sestu km 0,7, 90042 Monserrato (CA), Italy\\
$^{8}$Square Kilometre Array South Africa, Pinelands, 7405, South Africa\\
$^{9}$Department of Physics and Astronomy, West Virginia University, P.O. Box 6315, Morgantown, WV 26506, USA\\
$^{10}$Center for Gravitational Waves and Cosmology, Chestnut Ridge Research Building, Morgantown, WV 26505, USA
}}

\date{Accepted XXX. Received YYY; in original form ZZZ}

\pubyear{2015}

\begin{document}

\label{firstpage}
\pagerange{\pageref{firstpage}--\pageref{lastpage}}
\maketitle

\begin{abstract}
This paper is the second in a series where we report the results of the long-term timing of the millisecond pulsars (MSPs) in 47 Tucanae with the Parkes 64-m radio telescope. We obtain improved timing parameters that provide additional information for studies of the cluster dynamics: a) the pulsar proper motions yield an estimate of the proper motion of the cluster as a whole ($\mu_{\alpha}\, = \, 5.00\, \pm \, 0.14\, \rm mas \, yr^{-1}$, $\mu_{\delta}\, = \, -2.84\, \pm \, 0.12\, \rm mas \, yr^{-1}$) and the motion of the pulsars relative to each other. b) We measure the second spin-period derivatives caused by the change of the pulsar line-of-sight accelerations; 47~Tuc~H, U and possibly J are being affected by nearby objects. c) For ten binary systems we now measure changes in the orbital period caused by their acceleration in the gravitational field of the cluster. From all these measurements, we derive a cluster distance no smaller than $\sim\,$4.69 kpc and show that the characteristics of these MSPs are very similar to their counterparts in the Galactic disk. We find no evidence in favour of an intermediate mass black hole at the centre of the cluster. Finally, we describe the orbital behaviour of four  ``black widow'' systems. Two of them, 47~Tuc~J and O, exhibit orbital variability similar to that observed in other such systems, while for 47~Tuc~I and R the orbits seem to be remarkably stable. It appears, therefore, that not all ``black widows'' have unpredictable orbital behaviour.
\end{abstract}


%%%%%%%%%%%%%%%%%%%%%%%%%%%%%%%%%%%%%%%%%%%%%%%%%%
%%%%%%%%%%%%%%%%% BODY OF PAPER %%%%%%%%%%%%%%%%%%
\section{Introduction}
\label{sec:intro}

\begin{table*}
\label{tab:pulsars}
\begin{center}{\scriptsize
\setlength{\tabcolsep}{4pt}
\caption{References for the radio work on the pulsars in the globular cluster 47~Tucanae (NGC 104). Pulsars A, B and K do
not exist.
In boldface, we highlight the five pulsars discovered since 2000. 
The asterisks indicate the phase-coherent timing solutions determined since 2003.
In the second column, we describe the type of
system: WD binary implies that the companion is a white dwarf star, BW and RB imply black widow and
redback systems respectively, with an (e) indicating our detection of radio eclipses.
In the third column we indicate the reference of the
discovery, in the fourth column the reference with the first orbital solution precise enough to
predict the orbital phase for the whole data set. For instance, Camilo et al. (2000) published
approximate orbital parameters for pulsars P, R and W, but these were not precise enough to predict
the orbital phase many days in advance. All known binary systems now have well determined orbital
parameters. In the penultimate column we indicate the first publication with a timing solution
and in the last we list which timing solutions are presented in this work (and whether they are
updates or new solutions).}
\begin{tabular}{ l c c c c c}
\hline
Pulsar                                         &
\multicolumn{1}{c}{Type}             &
\multicolumn{1}{c}{Discovery}                  &
\multicolumn{1}{c}{Orbit}                      &
\multicolumn{1}{c}{First Timing Solution}                     &
\multicolumn{1}{c}{Timing}    
\\
47 Tuc                                       &
\multicolumn{1}{c}{}                           &
\multicolumn{1}{c}{(Reference)}                      &
\multicolumn{1}{c}{(Reference)}                      &
\multicolumn{1}{c}{(Reference)}                      &
\multicolumn{1}{c}{(This work)}                        
\\
\hline
C & isolated  & \cite{mlj+89a} & - & \cite{rlm+95}&  update \\
D & isolated  & \cite{mlr+91} & - & \cite{rlm+95}&  update \\
E & WD binary & \cite{mlr+91} & \cite{rlm+95} & \cite{fcl+01} & update \\
F & isolated  & \cite{mlr+91} & - & \cite{fcl+01} &  update \\
G & isolated  & \cite{mlr+91} & - & \cite{fcl+01} &  update \\
H & WD binary & \cite{mlr+91} & \cite{clf+00} & \cite{fcl+01} &  update \\ 
I & BW        & \cite{mlr+91} & \cite{rlm+95} & \cite{fcl+01} &  update \\
J & BW (e)    & \cite{mlr+91} & \cite{rlm+95} & \cite{clf+00} &  update \\
L & isolated  & \cite{mlr+91} & - & \cite{fcl+01} &  update \\
M & isolated  & \cite{mlr+91} & - & \cite{fcl+01} &  update \\
N & isolated  & \cite{rlm+95} & - & \cite{fcl+01} &  update \\
O & BW (e)    & \cite{clf+00} & \cite{clf+00} & \cite{fcl+01} &  update \\
P            & BW        & \cite{clf+00} & \cite{rfc+16} & \dots & \dots  \\
Q & WD binary & \cite{clf+00} & \cite{clf+00} & \cite{fcl+01} & update \\
R * & BW (e)    & \cite{clf+00} & This work & This work & new \\
S & WD binary & \cite{clf+00} & \cite{fkl01} & \cite{fck+03} &  update \\
T & WD binary & \cite{clf+00} & \cite{fkl01} & \cite{fcl+01} &  update \\
U & WD binary & \cite{clf+00} & \cite{clf+00} & \cite{fcl+01} &  update \\
V            & RB (e)    & \cite{clf+00} & \cite{rfc+16} & \dots  & \dots \\
W * & RB (e)    & \cite{clf+00} & \cite{rfc+16} & \cite{rfc+16} & no \\
{\bf X} * & WD binary & This work & \cite{rfc+16} & \cite{rfc+16} & no \\
{\bf Y} * & WD binary & This work & This work & This work & new \\
{\bf Z} * & isolated  & \cite{kni07a} & - & This work & new \\
{\bf aa} * & isolated & \cite{phl+16} & - & Freire \& Ridolfi (2017) &  no \\
{\bf ab} * & isolated & \cite{phl+16} & - &  \cite{phl+16} & update \\
\hline
\end{tabular}
}
\end{center}
\end{table*}

The year 2017 marks the $30^{\rm th}$ anniversary of the  discovery of the first radio
pulsar in a globular cluster (GC), PSR~B1821$-$24 \citep{lbm+87}. Since then
these clusters have been shown to be extremely prolific
millisecond pulsar (MSP) factories; the
total number of pulsars discovered in GCs is now\footnote{See
 \url{http://www.naic.edu/~pfreire/GCpsr.html} for an up-to-date list.}
 149, in a total of 28 Galactic globular
clusters (for a recent review, see \citealt{fre12}), the vast majority of which are MSPs. 
This makes the GC pulsar population very different from that of the Galactic disk.
The total number of radio pulsars in the Galactic GCs is probably of the order of
a few thousand \citep{tl13}, most of which have not yet been found because of the 
insufficient sensitivity and sky coverage of extant radio telescopes. Their discovery will
probably have to wait for the construction of the Five hundred meter Aperture
Spherical Telescope \citep{slk+09}
or the Square Kilometer Array \citep[SKA,][]{hpb+15}.

The globular cluster 47 Tucanae (also known as NGC 104, henceforth 47~Tuc) has a total of
25 known radio pulsars, second only to the GC Terzan 5, which has 37 known pulsars
\citep[see e.g.,][Cadelano et al. in prep.]{rhs+05,hrs+06}. All pulsars in 47~Tuc have spin periods smaller
than 8 ms; of these, 15 are in binary systems (see Table \ref{tab:pulsars}). 
These discoveries and subsequent timing (see also Table \ref{tab:pulsars} for the references)
have enabled unprecedented studies of stellar evolution in GCs
\citep{rpr00}, studies of cluster dynamics \citep{fck+03} and even the discovery
of ionised gas in the cluster, the first ever detection of any sort of interstellar medium within a GC \citep{fkl+01}. In addition, all the pulsars with well-determined
positions have been identified at X-ray wavelengths
\citep[][Bhattacharya et al., submitted]{ghem01,gch+02,hge+05,bgh+06,rfc+16} and at least 6
companion objects have been identified at optical
wavelengths \citep{egh+01,egc+02,rbh+15,cpf+15}.

In the first paper from this series \citep[henceforth Paper I]{rfc+16}
we described the motivation, observations and data processing
of the long-term radio monitoring of the radio pulsars in 47 Tuc with the
64-m Parkes radio telescope.
That paper focused on one of the objectives of the long-term timing,
the characterization of the elusive binary pulsars 47~Tuc~P, V, W and X;
for two of those systems (47~Tuc~W and X) it was possible to
derive phase-coherent timing solutions. Of these systems,
47~Tuc~X is especially interesting -- it is a binary with an extremely low eccentricity
that lies well outside the central regions of the cluster.

In this paper, the second in the series, we present up-to-date timing solutions for
20 of the 25 MSPs in 47~Tuc, which include all data from the Analogue Filterbank
(AFB, see Section~\ref{sec:data}). The bulk of the paper is a discussion of some of
the implications of the 23 known timing solutions (we
also use the solutions for 47~Tuc~W and X derived in Paper I,
and the timing solution for 47~Tuc~aa, to be presented in Freire \& Ridolfi 2017, in prep.). 

We discuss the proper motions in
Section~\ref{sec:proper_motions}, the new ``jerk'' measurements in
Section~\ref{sec:jerks} and in Section~\ref{sec:accelerations}
we discuss the variation of the orbital periods for a set of ten binary systems. This includes
estimates of the line-of-sight component of the accelerations of the systems in the gravitational
field of the cluster, which can be used to determine their true spin-down parameters.
In Section~\ref{sec:spiders}, we discuss the long-term orbital behaviour of the four
``black widow" systems\footnote{See \citet{Roberts13} for a review of ``black widow''
and ``redback'' pulsar binary systems} with known timing solutions in 47 Tuc, and in
Section~\ref{sec:binary_masses} we present the detection of the rate of advance of
periastron in three MSP-WD binaries and a refined measurement of the mass of the
47~Tuc~H binary system. We discuss some of the implications of our results
in Section~\ref{sec:discussion} and summarise our findings in Section~\ref{sec:conclusions}.

All of these results benefit greatly from the much
larger timing baseline presented in this paper in comparison with that in previous publications.
They provide information that will be used to produce improved dynamical models of the cluster,
a better model of the intracluster gas distribution, and will provide input for stellar
evolution models, all of these being strong motivations for the long-term timing of the cluster.

\begin{figure}
    \includegraphics[width=1\columnwidth]{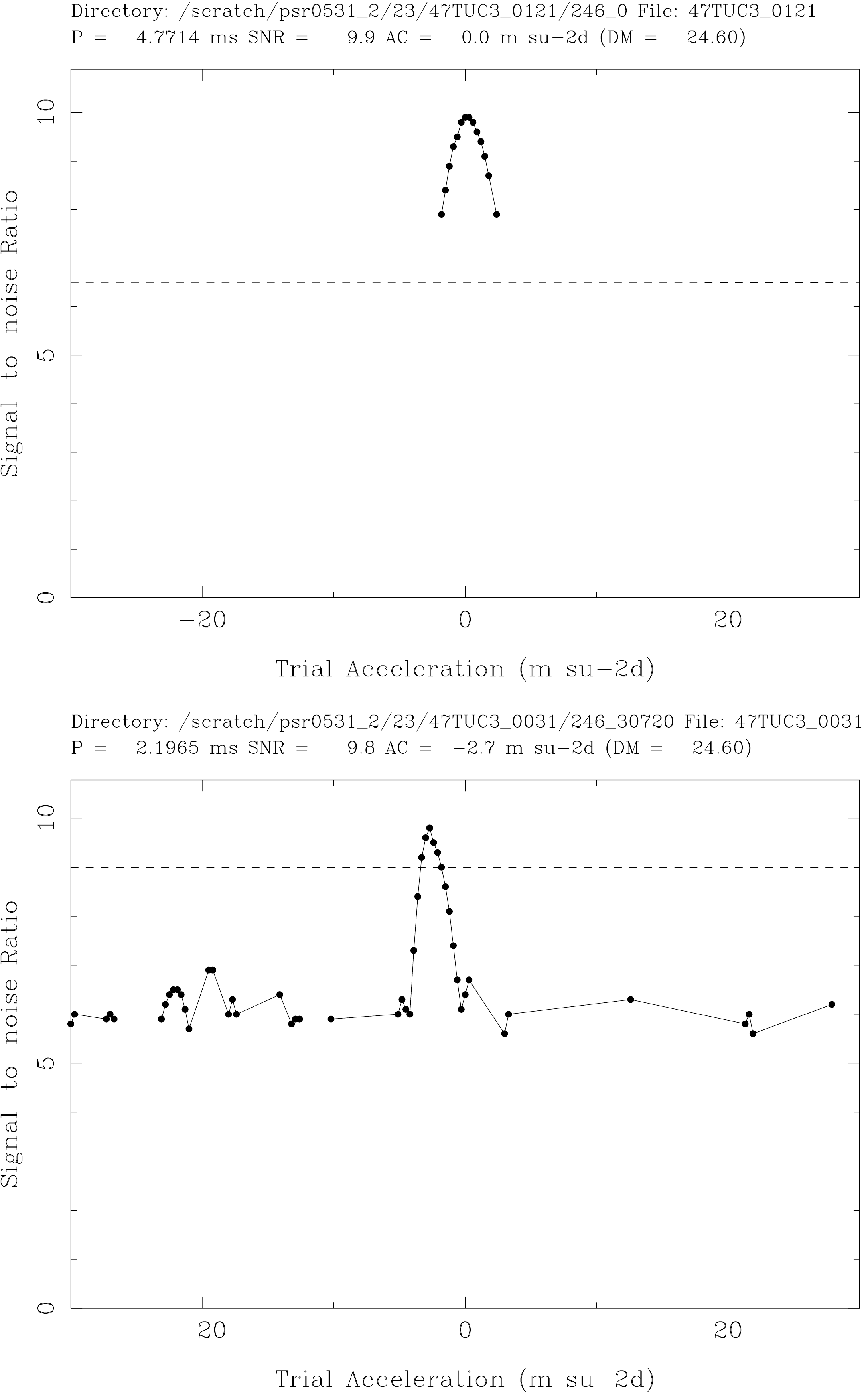}
    \caption{Discovery plots for 47~Tuc~X (top) and 47~Tuc~Y (bottom)
    from the AS23 survey, described by \citet{clf+00}. Each plot shows how the
    signal-to-noise ratio varies as a function of trial acceleration. These
    candidates could not be confirmed within the scope of AS23; they were later
    confirmed by the AS25 survey (see text).}
    \label{fig:new_pulsars}
\end{figure}

\begin{figure*}
    \includegraphics[width=0.95\textwidth]{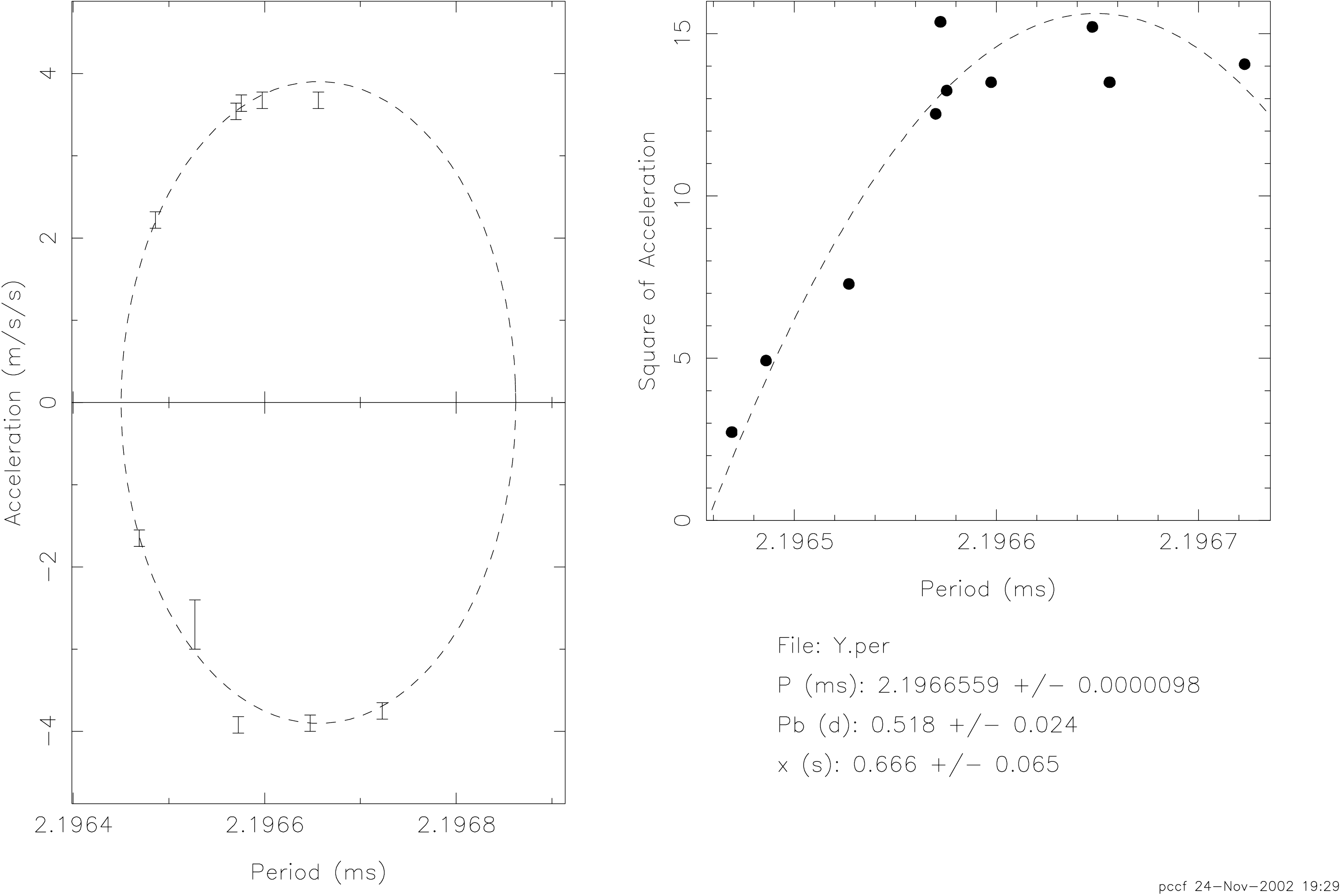}
    \caption{For 47~Tuc~Y, the AS25 detections were
    sufficient for the determination of the orbit using the {\sc circorbit}
    routine, described by \citet{fkl01}.
    In the right panel, a parabola was fit to the squares of the acceleration
    as a function of spin period. This is translated (in the left panel)
    to a best-fitting ellipse (dashed) to the
    observed accelerations as a function of barycentric spin period, represented
    by the vertical error bars.
    This ellipse corresponds to the spin and orbital parameters listed in the
    bottom right.}
    \label{fig:Y}
\end{figure*}

\subsection{Cluster parameters}

In the analysis presented below, we benefit greatly from the new studies of 
{\em Hubble Space Telescope} (\emph{HST}) data that have become available since
our last study of the cluster potential \citep{fck+03},
these provide much more precise cluster parameters and in some cases
entirely new information. \cite{mam+06}
placed the centre of 47~Tuc at right ascension
$\alpha\, = \, 00^{\rm h}\, 24^{\rm m}\, 5\fs67$ and declination
$\delta\, = \, -72^\circ\, 04\arcmin\, 52\farcs62$; they also
measured the angular core radius: $\theta_c \, = \, 0.347\, \arcmin$.
A newly available measurement, which will be of great importance
for this work, is the 1-D proper motion dispersion
for stars at the cluster centre obtained from differential
\emph{HST} astrometry: $\sigma_{\mu, 0}\, = \, 0.573 \, \rm mas\, yr^{-1}$ \citep{wmba15}.

From Eq. 1-34 in \cite{spi87}, which is accurate to $\sim$0.5\%
for clusters where the tidal radius
$r_t$ is much larger than the core radius $r_c$
(such as 47~Tuc, where $c = \log_{10} (r_t / r_c) = 2.07$, see \citealt{har96}),
we derive the following expression for the central density:
\begin{equation}
\label{eq:central_density}
\rho(0) \, = \, \frac{9 \sigma^2_{\mu, 0}}{4 \pi G \theta_c^2} 
\end{equation}
where we have replaced the spectroscopic radial velocity (RV)
dispersion $\sigma_0$ with $\sigma_{\mu, 0} \, d$
(where $d$ is the distance to 47 Tuc)
and the core radius $r_c$ with $\theta_c d$. The
distance terms then cancel out; this means that the central density
can be determined solely from the aforementioned angular measurements,
independently of $d$. For the $\sigma_{\mu, 0}$ and $\theta_c$
of 47~Tuc Eq.~\ref{eq:central_density} yields
$\rho(0) \, = \, 1.20 \times 10^5 M_{\odot} \rm pc^{-3}$.

For $d$, we use 4.69 kpc \citep{wgk+12}. Other recent assessments
place the cluster at very similar distances; for instance,
using the (relatively well-trusted) white dwarf (WD) cooling track model,
\cite{hka+13} derived $d\, =\, 4.6 \pm 0.2$ kpc. Using other methods,
like the self-consistent isochrone fits to
colour-magnitude diagrams and the eclipsing binary star V69,
\cite{bbb+17} derived a slightly smaller $d\, =\, 4.4 \pm 0.2$ kpc.
Averaging several recent measurements, \cite{bhog16} obtained
$d\, =\,4.53^{+0.08}_{-0.04}$ kpc.

However, not all distance estimates match:
By comparing their measurement of $\sigma_{\mu, 0}$ to their best estimate of
$\sigma_0$, \cite{wmba15b} derived a kinematic 
$d\, =\,4.15 \, \pm \, 0.08\, \rm kpc$,
consistent with the earlier estimate of \cite{mam+06},
$d \, = \, 4.02\, \pm \, 0.35 \, \rm kpc$.

A possible explanation of this discrepancy is that the
$\sigma_{0}$ measurements (generally close to 11 km s$^{-1}$), are biased towards smaller values.
Likely reasons for this were discussed in detail in \cite{bhog16}.
Briefly, those authors pointed out that the RV
measurements used for comparison were intended to be of single stars,
however \emph{HST} images show that a number of the targeted ``stars''
actually comprise more than one star of similar brightness. The RV 
measurements of combined stars tend to be closer to the cluster mean
than single stars, so this has the effect of reducing the inferred
velocity dispersion and the resultant $d$.
Indeed, when \cite{wmba15b} included a larger sample of RV
measurements that extend further out from the core of 47 Tuc to
compare with their proper motions, they found
$d\, =\,4.61^{+0.06}_{-0.07}\, $kpc (see their Appendix A, Fig. 9),
consistent with the larger distances mentioned above.

This issue is crucial for the interpretation of
our results. As we will see in
Sections~\ref{sec:jerks} and \ref{sec:accelerations}, our
results also favour this large distance, rather than the smaller
kinematic distance estimates. This issue is also crucial, as we shall
see, for addressing the question of the presence of an intermediate
mass black hole in the centre of 47~Tucanae, which has been
repeatedly raised in the literature.

\section{Data reduction}
\label{sec:data}

The data analysed here were taken with the Australian 64-m Parkes radio telescope.
The observations, receivers and signal-processing systems were described in Paper I. Two filterbank systems
were used: the {\em low}-resolution $2\, \times \, 96 \, \times\, 3$ MHz analogue filterbank
and the $2\, \times\, 512\, \times \, 0.5$ MHz {\em high}-resolution analogue filterbanks, henceforth
the lAFB and hAFB, respectively.

\subsection{Discovery of two millisecond pulsars}

\begin{figure*}
    \includegraphics[width=0.9\textwidth]{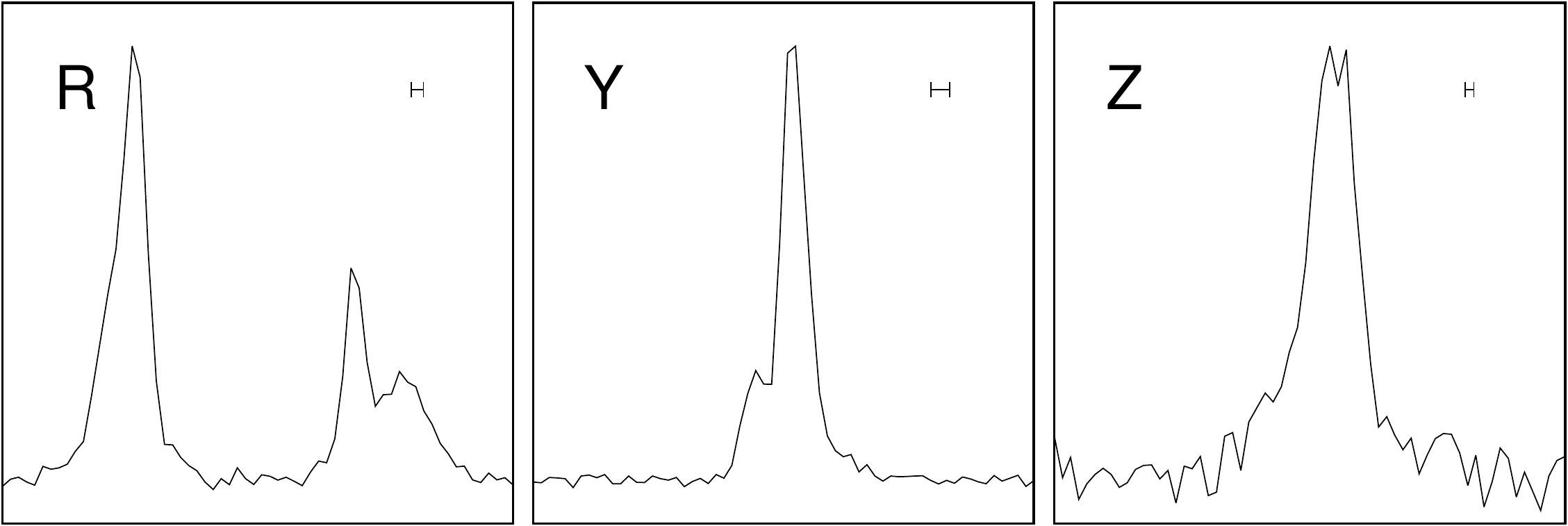}
    \caption{Full-cycle pulse profiles of 47 Tuc R, Y and Z. The horizontal error bars display
    the time resolution of the hAFB data.}
    \label{fig:profiles_RYZ}
\end{figure*}

Before moving to the bulk of the paper, we report on the discovery of two MSPs,
47 Tuc~X and Y. These pulsars, with periods of 4.771 and 2.196 ms (see Fig.~\ref{fig:new_pulsars})
were initially found in the survey described
by \cite{clf+00}, but could only be confirmed in a deeper 
survey briefly described in \cite{fkl01}.
This used the same software (the {\sc SIGPROC} acceleration search 
routines\footnote{\url{http://sigproc.sourceforge.net/}}) and data (the lAFB data) used in \cite{clf+00}.
The only difference was that instead of searching 17.5-minute segments of data, each containing
$2^{23}$ 125-$\mu$s-long time samples,
it used 70-minute segments, each containing $2^{25}$ time samples;
for this reason we designate the former AS23 (where AS means ``accelerated search'')
and the latter AS25.
Because of computing limitations, AS25 only covered accelerations from $-5$ to $5\, \rm m \, s^{-2}$
in steps of $0.02\, \rm m \, s^{-2}$, while AS23 covered accelerations from
$-30$ to $30\, \rm m \, s^{-2}$ in steps of $0.3\, \rm m \, s^{-2}$.
In the case of sources with steady flux density,
AS25 has twice the sensitivity of AS23 for pulsars with small accelerations.

Some of the earlier results of AS25 were reported in \cite{fkl01}: 47~Tuc~T and S (which
are always within its acceleration range) were detected enough times to allow
the determination of their orbits using the new orbital determination
method presented in that paper. Later, 47~Tuc~X and Y were confirmed by
this survey. These two
pulsars were also independently discovered in a parallel search by one of
us, N. D'Amico, in Bologna, Italy. Their existence was mentioned
for the first time in \cite{lcf+03}.

\subsection{Determination of orbits and timing solutions}
\label{sec:timing}

For 47~Tuc~Y, the AS25 detections were
numerous enough to determine the orbit using the period - acceleration
method of \citet{fkl01} (see Fig.~\ref{fig:Y}). The orbit
is nearly circular and has a period of
12.52 hours and, assuming a pulsar mass of 1.4 M$_{\odot}$,
a minimum companion mass of 0.141 M$_{\odot}$. 
Its parameters were then refined in three stages:
1) finding the correct orbit count using the
method described in Paper I (i.e., searching for an orbital period
that is the common integer sub-multiple of all differences
between times of passage through ascending node),
2) by fitting an orbital model to the variation of the spin period versus time,
and 3) fitting pulse times of arrival (ToAs) as a function of time using the
{\sc TEMPO} pulsar timing 
software\footnote{\url{http://tempo.sourceforge.net/}};
the ToAs for this pulsar were derived with the pulse profile presented in 
Fig~\ref{fig:profiles_RYZ}.
For 47~Tuc~X, a more sophisticated and thorough search of the
hAFB data was necessary for determining its orbit and is reported in Paper I,
its pulse profile is also presented there.

After 2003, the AS23 and AS25 searches were extended to all the
hAFB data then available. This resulted in two extra detections
of 47~Tuc~R, an eclipsing black widow with a minimum companion mass of
only 0.0264 M$_{\odot}$ (assuming $M_p \, = \, 1.4 \, \rm M_{\odot}$)
and at the time the binary pulsar with the shortest orbital
period known \citep[96 minutes,][]{clf+00}.
We then determined a precise orbital period
for this system using (again) the orbital count technique
described in Paper I. Its pulse profile, based on hAFB data,
is displayed in  Fig.~\ref{fig:profiles_RYZ}. This represents a substantial
improvement over the previous profile based on lAFB data in \cite{clf+00}.

More recently, and using the full hAFB data set,
we have been able to detect the isolated MSP 47~Tuc~Z (discovered
by \citealt{kni07a}) enough times to derive a good preliminary ephemeris.
Its pulse profile is also displayed in Fig.~\ref{fig:profiles_RYZ}.

For all these pulsars (R, Y and Z), the subsequent processing was the same.
All data were folded using their preliminary ephemerides. This
increased the number of detections greatly and led to the determination
of even better ephemerides, which in turn allowed
even more detections. This iterative process eventually 
allowed the determination of phase-coherent timing solutions of these pulsars, which are
presented here for the first time.

Because the timing solutions of 47~Tuc~R and Y were well determined in 2006,
both were included in the group of 19 MSPs for which X-ray emission
was detected with the {\em Chandra} X-ray observatory in \cite{bgh+06}.
For 47~Tuc~Z, aa and ab, the timing solutions
and resulting precise positions were only determined in 2016; their {\em Chandra}
X-ray analysis is presented in Bhattacharya et al. 2017 (submitted). %C. Heinke
The pulsars 47~Tuc~R and Y have also been well studied at optical wavelengths.
The WD companion of 47~Tuc~Y is clearly detected, and it is the second
brightest WD after the WD companion to 47~Tuc~U \citep{rbh+15,cpf+15}.

\subsection{Updated timing solutions}

For this work, all the extant AFB data were de-dispersed and folded anew using the 
{\sc DSPSR} routine \citep{van_Straten_Bailes11} with the best previous ephemerides
for 20 MSPs. The resulting pulse profiles were then cross-correlated with a low-noise 
profile, normally derived from the best detection(s) of each pulsar, using the method 
described in \cite{tay92} and implemented in the {\sc PSRCHIVE} software 
\citep{Hotan+04,van_Straten+12}.

This resulted in topocentric ToAs which were then analysed with {\sc TEMPO},
where we used the Jet Propulsion Laboratory (JPL) DE/LE 421 Solar system
ephemeris \citep{Folkner+2009} to subtract the effect of the motion of the
radio telescope relative to the barycentre of the Solar System.

The timing solutions of the isolated pulsars are presented in Table~\ref{tab:isolated},
those of the MSP-WD systems in Table~\ref{tab:MSP_WD} and the solutions of
four black widow systems in Table~\ref{tab:MSP_BW}.
The {\sc TEMPO} ephemerides are available online\footnote{\url{http://www3.mpifr-bonn.mpg.de/staff/pfreire/47Tuc/}}.
These solutions describe the data well, with no trends detectable in the ToA residuals,
except for slight delays near eclipse phase for some of the eclipsing pulsars
\citep{fck+03}; these are not taken into account in the solutions. The uncertainties 
presented in all these tables are 1-$\sigma$ (68\%) confidence limits, and were derived 
using a Monte-Carlo bootstrap routine implemented in {\sc TEMPO}.  In all solutions,
we fixed a parallax that corresponds to the assumed distance to 47~Tuc,
4.69 kpc \citep{wgk+12}.

%%%%%%%%%%%%%%%%%%%%%%%%%%%%%%%%%%%%%%%%%%%%%%%%%%%%%%%%%%%%%%%
% TABLE OF ISOLATED PULSARS
%%%%%%%%%%%%%%%%%%%%%%%%%%%%%%%%%%%%%%%%%%%%%%%%%%%%%%%%%%%%%%%

\begin{table*}
\caption{Timing parameters for nine of the ten isolated pulsars in 47 Tuc, as obtained from fitting the observed ToAs with {\sc TEMPO}. For all the pulsars in this and the following tables, the reference epoch is MJD 51600; the 1-$\sigma$ uncertainties are calculated using a Monte-Carlo bootstrap routine implemented in {\sc TEMPO}. A fixed parallax value of 0.2132 mas was assumed; the time units are TDB; the adopted terrestrial time standard is UTC(NIST); the Solar System ephemeris used is JPL DE421. }
\label{tab:isolated}
\begin{center}{\scriptsize
\setlength{\tabcolsep}{4pt}
\renewcommand{\arraystretch}{1.3}
\begin{tabular}{l c c c c}
\hline
Pulsar  &   47 Tuc C                                                           &   47 Tuc D                                                         &    47 Tuc F                                                           &  47 Tuc G                                                             \\
\hline\hline
Right Ascension, $\alpha$ (J2000)                                     \dotfill &   00:23:50.3546(1)                                                       &   00:24:13.88092(6)                                                      &   00:24:03.85547(10)                                                     &   00:24:07.9603(1)                                                       \\
Declination, $\delta$ (J2000)                                         \dotfill &   $-$72:04:31.5048(4)                                                      &   $-$72:04:43.8524(2)                                                      &   $-$72:04:42.8183(2)                                                      &   $-$72:04:39.7030(5)                                                      \\
Proper Motion in $\alpha$, $\mu_\alpha$ (mas yr$^{-1}$)               \dotfill &   5.2(1)                                                                 &   4.24(7)                                                                &   4.52(8)                                                                &   4.5(1)                                                                 \\
Proper Motion in $\delta$, $\mu_\delta$ (mas yr$^{-1}$)               \dotfill &   $-$3.1(1)                                                                &   $-$2.24(5)                                                               &   $-$2.50(5)                                                               &   $-$2.9(1)                                                                \\
%Parallax (mas)                                                        \dotfill &   0.2132                                                                 &   0.2132                                                                 &   0.2132                                                                 &   0.2132                                                                 \\
Spin Frequency, $f$ (Hz)                                        \dotfill &   173.708218965958(4)                                                    &   186.651669856731(3)                                                    &   381.158663656311(5)                                                    &   247.501525096385(8)                                                    \\
First Spin Frequency derivative, $\dot{f}$ ($10^{-15}$ Hz s$^{-1}$)                \dotfill &   1.50421(6)    &   0.11922(3)   &   $-$9.3711(1) &   2.5825(1) \\
Second Spin Frequency derivative, $\ddot{f}$ ($10^{-27}$ Hz s$^{-2}$)               \dotfill &   1.3(4) &   $-$1.2(2) &   6.8(7) &   6.0(9) \\
%Reference Epoch (MJD)                                                 \dotfill &   51600.000000                                                           &   51600.000000                                                           &   51600.000000                                                           &   51600.000000                                                           \\
Start of Timing Data (MJD)                                            \dotfill &   47857.439                                                              &   47716.842                                                              &   48494.790                                                              &   48600.489                                                              \\
End of Timing Data (MJD)                                              \dotfill &   56508.971                                                              &   56508.976                                                              &   56466.879                                                              &   56466.879                                                              \\
Dispersion Measure, DM (pc cm$^{-3}$)                                 \dotfill &   24.600(4)                                                              &   24.732(3)                                                              &   24.382(5)                                                              &   24.436(4)                                                              \\
%Solar System Ephemeris                                                \dotfill &   DE421                                                                  &   DE421                                                                  &   DE421                                                                  &   DE421                                                                  \\
%Terrestrial Time Standard                                             \dotfill &   UTC(NIST)                                                              &   UTC(NIST)                                                              &   UTC(NIST)                                                              &   UTC(NIST)                                                              \\
%Time Units                                                            \dotfill &   TDB                                                                    &   TDB                                                                    &   TDB                                                                    &   TDB                                                                    \\
Number of ToAs                                                        \dotfill &   6225                                                                   &   3607                                                                   &   1785                                                                   &   594                                                                    \\
Residuals RMS ($\mu$s)                                                \dotfill &   12.33                                                                  &   8.74                                                                   &   7.83                                                                   &   11.25                                                                  \\
\hline
\multicolumn{5}{c}{Derived Parameters}  \\
\hline
Angular offset from centre in $\alpha$, $\theta_{\alpha}$ (arcmin) \dotfill & $-$1.1784 & +0.6316 & $-$0.1396 & +0.1762 \\
Angular offset from centre in $\delta$, $\theta_{\delta}$ (arcmin) \dotfill & +0.3520   & +0.1460 & +0.1634 & +0.2151 \\
Total angular offset from centre, $\theta_{\perp}$ (arcmin) \dotfill & 1.2298 & 0.6483 & 0.2149 & 0.2781 \\
Total angular offset from centre, $\theta_{\perp}$ (core radii) \dotfill & 3.5442 & 1.8683 & 0.6194 & 0.8014 \\
Projected distance from centre, $r_{\perp}$ (pc) \dotfill & 1.6778 & 0.8845 & 0.2932 & 0.3794 \\
Spin Period, $P$ (ms)                                                  \dotfill &   5.7567799955164(1)                                     &   5.35757328486572(7)                                    &   2.62357935251262(3)                                    &   4.0403791435651(1)                                     \\
First Spin Period derivative, $\dot{P}_{\rm obs}$ ($10^{-21}$ s s$^{-1}$)                    \dotfill &   $-$49.850(2)                                            &   $-$3.4219(9)                                            &   64.5031(7)                                            &   $-$42.159(2)                                            \\
Line-of-sight jerk, $\dot{a}_\ell$ ($10^{-21}\, \rm m\, s^{-3}$) \dotfill & $-$2.3(7) & 1.98(26) & $-$5.3(5) & $-$7.3(1.1) \\
\hline
\end{tabular} }
\end{center} 
%\end{table*}
%\begin{table*}
\contcaption{}
\label{tab:isolated_continued}
\begin{center}{\scriptsize
\setlength{\tabcolsep}{4pt}
\renewcommand{\arraystretch}{1.3}
 \begin{tabular}{l c c c c c}
\hline
Pulsar &  47 Tuc L                                                            &  47 Tuc M                                                            &  47 Tuc N                                                            &  47 Tuc Z                                                            &  47 Tuc ab                                                           \\
\hline\hline

Right Ascension, $\alpha$ (J2000)                                     \dotfill &   00:24:03.7721(3)                                                       &   00:23:54.4899(3)                                                       &   00:24:09.1880(2)                                                       &   00:24:06.041(2)                                                        &   00:24:08.1615(5)                                                       \\
Declination, $\delta$ (J2000)                                         \dotfill &   $-$72:04:56.923(2)                                                     &   $-$72:05:30.756(2)                                                       &   $-$72:04:28.8907(7)                                                      &   $-$72:05:01.480(6)                                                       &   $-$72:04:47.602(2)                                                       \\
Proper Motion in $\alpha$, $\mu_\alpha$ (mas yr$^{-1}$)               \dotfill &   4.4(2)                                                                 &   5.0(3)                                                                 &   6.3(2)                                                                 &   4(2)                                                                   &   4.2(6)                                                                 \\
Proper Motion in $\delta$, $\mu_\delta$ (mas yr$^{-1}$)               \dotfill & $-$2.4(2)                                                                  &   $-$2.0(4)                                                                &   $-$2.8(2)                                                                &   1(2)                                                                   &   $-$2.9(5)                                                                \\
%Parallax (mas)                                                        \dotfill &   0.2132                                                                 &   0.2132                                                                 &   0.2132                                                                 &   0.2132                                                                 &   0.2132                                                                 \\
Spin Frequency, $f$ (Hz)                                        \dotfill &  230.08774629142(2)                                                     &   271.98722878874(2)                                                     &   327.44431861739(1)                                                     &    219.5656060346(1)                                                    &   269.93179806134(4)                                                     \\
First Spin Frequency derivative, $\dot{f}$ ($10^{-15}$ Hz s$^{-1}$)                \dotfill &   6.4611(2) &   2.8421(4) &   2.3435(2) &  0.219(3) & $-$0.7155(6) \\
Second Spin Frequency derivative, $\ddot{f}$ ($10^{-27}$ Hz s$^{-2}$)  \dotfill &   $-$1.3(1.3) &   7(2) &   $-$9(2) &     8(25) &   $-$8(3) \\
%Reference Epoch (MJD)                                                 \dotfill &   51600.000000                                                           &   51600.000000                                                           &   51600.000000                                                           &   51600.000000                                                           &   51600.000000                                                           \\
Start of Timing Data (MJD)                                            \dotfill &   50686.683                                                              &   48491.694                                                              &   48515.534                                                              &   51003.792                                                              &   51000.785                                                              \\
End of Timing Data (MJD)                                              \dotfill &   56388.208                                                              &   55526.513                                                              &   55648.110                                                              &   54645.852                                                              &   56388.135                                                              \\
Dispersion Measure, DM (pc cm$^{-3}$)                                 \dotfill &   24.40(1)                                                              &   24.43(2)                                                               &   24.574(9)                                                              &   24.45(4)                                                              &   24.37(2)                                                              \\
%Solar System Ephemeris                                                \dotfill &   DE421                                                                  &   DE421                                                                  &   DE421                                                                  &   DE421                                                                  &   DE421                                                                  \\
%Terrestrial Time Standard                                             \dotfill &   UTC(NIST)                                                              &   UTC(NIST)                                                              &   UTC(NIST)                                                              &   UTC(NIST)                                                              &   UTC(NIST)                                                              \\
%Time Units                                                            \dotfill &   TDB                                                                    &   TDB                                                                    &   TDB                                                                    &   TDB                                                                    &   TDB                                                                    \\
Number of ToAs                                                        \dotfill &   411                                                                    &   315                                                                    &   436                                                                    &   107                                                                    &   210                                                                    \\
Residuals RMS ($\mu$s)                                                \dotfill &   17.02                                                                  &   20.15                                                                  &   12.98                                                                  &   58.78                                                                  &   24.85                                                                  \\
\hline
\multicolumn{6}{c}{Derived Parameters}  \\
\hline
Angular offset from centre in $\alpha$, $\theta_{\alpha}$ (arcmin) \dotfill & $-$0.1460 & $-$0.8594 & +0.2707 & +0.0286 & +0.1917 \\
Angular offset from centre in $\delta$, $\theta_{\delta}$ (arcmin) \dotfill & $-$0.0719 & $-$0.6354 & +0.3955 & $-$0.1479 & +0.0838 \\
Total angular offset from centre, $\theta_{\perp}$ (arcmin) \dotfill & 0.1627 & 1.0688 & 0.4793 & 0.1506 & 0.2092 \\
Total angular offset from centre, $\theta_{\perp}$ (core radii) \dotfill & 0.4689 & 3.0801 & 1.3812 & 0.4340 & 0.6028 \\
Projected distance from centre, $r_{\perp}$ (pc) \dotfill & 0.2220 & 1.4581 & 0.6539 & 0.2054 & 0.2854 \\
Spin Period, $P$ (ms)                                                  \dotfill &   4.3461679994616(3)                                     &   3.6766432176002(3)                                     &   3.0539543462608(1)                                    &   4.554447383906(3)                                      &   3.7046394947985(5)                                    \\
First Spin Period derivative, $\dot{P}_{\rm obs}$ ($10^{-21}$ s s$^{-1}$)                    \dotfill &   $-$122.0406(10)                                           &   $-$38.418(5)                                            &   $-$21.857(2)                                            &   $-$4.56(1)                                              &   9.820(8)                                              \\
Line-of-sight jerk, $\dot{a}_\ell$ ($10^{-21}\, \rm m\, s^{-3}$) \dotfill & 1.7(1.7) & $-$8.0(2.6) & 8.5(1.5) & $-$11(33) & 8.7(3.6) \\
\hline
\end{tabular} }
\end{center}
\end{table*}

%%%%%%%%%%%%%%%%%%%%%%%%%%%%%%%%%%%%%%%%%%%%%%%%%%%%%%%%%%%%%%%
% TABLE OF WD-pulsar BINARIES
%%%%%%%%%%%%%%%%%%%%%%%%%%%%%%%%%%%%%%%%%%%%%%%%%%%%%%%%%%%%%%%

\begin{table*}
\caption{Timing parameters for seven of the eight MSP-WD binaries in 47 Tuc, as obtained from fitting the observed ToAs with {\sc TEMPO}. The eighth MSP-WD system, 47~Tuc~X, was already studied in detail in Paper I. The orbital models used are DD \citep{dd85,dd86} and ELL1 \citep{lcw+01}.
For the characteristic age, we either present the median or a 2-$\sigma$ lower limit.}
\label{tab:MSP_WD} 
\begin{center}{\scriptsize
\setlength{\tabcolsep}{6pt}
\renewcommand{\arraystretch}{1.3}
\begin{tabular}{l c c c c}
\hline
Pulsar &  47 Tuc E & 47 Tuc H & 47 Tuc Q & 47 Tuc S  \\
\hline\hline

Right Ascension, $\alpha$ (J2000)                                     \dotfill &   00:24:11.10528(5)                                                      &   00:24:06.7032(2)                                                       &   00:24:16.4909(1)                                                       &   00:24:03.9794(1)                                                       \\
Declination, $\delta$ (J2000)                                         \dotfill &   $-$72:05:20.1492(2)                                                      &   $-$72:04:06.8067(6)                                                      &   $-$72:04:25.1644(6)                                                      &   $-$72:04:42.3530(4)                                                      \\
Proper Motion in $\alpha$, $\mu_\alpha$ (mas yr$^{-1}$)               \dotfill &   6.15(3)                                                                &   5.1(2)                                                                 &   5.2(1)                                                                 &   4.5(1)                                                                 \\
Proper Motion in $\delta$, $\mu_\delta$ (mas yr$^{-1}$)               \dotfill &   $-$2.35(6)                                                               &   $-$2.8(2)                                                                &   $-$2.6(1)                                                                &   $-$2.5(1)                                                                \\
% Parallax (mas)                                                        \dotfill &   0.2132                                                                 &   0.2132                                                                 &   0.2132                                                                 &   0.2132                                                                 \\
Spin Frequency, $f$ (Hz)                                        \dotfill &   282.779107035000(3)                                                    &      311.49341784423(1)                                                  &     247.943237418920(9)                                                  &   353.306209385356(9)                                                    \\
First Spin Frequency derivative, $\dot{f}$ ($10^{-15}$ Hz s$^{-1}$)                \dotfill &   $-$7.87728(4) &   0.1775(1)  &   $-$2.0907(2) &   15.0466(1) \\
Second Spin Frequency derivative, $\ddot{f}$ ($10^{-27}$ Hz s$^{-2}$)   \dotfill &   2.9(2)  &   1.60(2)$\times 10^{-25}$                                              &      7(11)$\times 10^{-28}$                                                                  &   -7.8(8)$\times 10^{-27}$                                               \\
%Reference Epoch (MJD)                                                 \dotfill &   51600.000000                                                           &   51600.000000                                                           &   51600.000000                                                           &   51600.000000                                                           \\
Start of Timing Data (MJD)                                            \dotfill &   48464.854                                                              &   48517.512                                                              &   50689.700                                                              &   50686.683                                                              \\
End of Timing Data (MJD)                                              \dotfill &   56508.948                                                              &   56508.972                                                              &   56388.178                                                              &   56466.879                                                              \\
Dispersion Measure, DM (pc cm$^{-3}$)                                 \dotfill &   24.236(2)                                                              &   24.369(8)                                                              &   24.265(4)                                                              &   24.376(4)                                                              \\
%Solar System Ephemeris                                                \dotfill &   DE421                                                                  &   DE421                                                                  &   DE421                                                                  &   DE421                                                                  \\
%Terrestrial Time Standard                                             \dotfill &   UTC(NIST)                                                              &   UTC(NIST)                                                              &   UTC(NIST)                                                              &   UTC(NIST)                                                              \\
%Time Units                                                            \dotfill &   TDB                                                                    &   TDB                                                                    &   TDB                                                                    &   TDB                                                                    \\
Number of ToAs                                                        \dotfill &   1812                                                                   &   1073                                                                   &   697                                                                    &   577                                                                    \\
Residuals RMS ($\mu$s)                                                \dotfill &   6.06                                                                   &   17.04                                                                  &   12.73                                                                  &   9.50                                                                   \\

\hline
\multicolumn{5}{c}{Binary Parameters}  \\
\hline
 Binary Model                                                          \dotfill &   DD                                                                     &   DD                                                                     &   ELL1                                                                   &   ELL1                                                                   \\
Projected Semi-major Axis, $x_p$ (lt-s)                               \dotfill &   1.9818427(4)                                                           &   2.152813(2)                                                            &   1.4622043(9)                                                           &   0.7662686(8)                                                           \\
Orbital Eccentricity, $e$                                             \dotfill &   3.159(4)$\times 10^{-4}$                                               &   7.0558(1)$\times 10^{-2}$                                              &   --                                                                     &   --                                                                     \\
Longitude of Periastron, $\omega$ (deg)                               \dotfill &   218.6(1)                                                               &   110.603(1)                                                             &   --                                                                     &   --                                                                     \\
Epoch of passage at Periastron, $T_0$ (MJD)                           \dotfill &   51001.7900(8)                                                          &   51602.186289(7)                                                        &   --                                                                     &   --                                                                     \\
First Laplace-Lagrange parameter, $\eta$                              \dotfill &   --                                                                     &   --                                                                     &   6.2(1)$\times 10^{-5}$                                                 &   9.1(3)$\times 10^{-5}$                                                 \\
Second Laplace-Lagrange parameter, $\kappa$                            \dotfill &   --                                                                     &   --                                                                     &   $-$5.1(2)$\times 10^{-5}$                                                &   3.87(2)$\times 10^{-4}$                                                \\
Epoch of passage at Ascending Node, $T_\textrm{asc}$ (MJD)            \dotfill &   --                                                                     &   --                                                                     &    51600.2842078(2)                                                     &   51600.6250241(2)                                                       \\
Rate of periastron advance, $\dot{\omega}$ (deg/yr)                   \dotfill &   0.090(16)                                                   &   0.06725(19)                                                &   --                                                                     &   0.331(75)                                                                \\
Orbital Period, $P_{\rm b}$ (days)                                          \dotfill &   2.2568483(9)                                                           &    2.357696895(10)                                                       &    1.1890840496(4)                                                       &   1.2017242354(6)                                                        \\
Orbital Period derivative, $\dot{P}_b$ (10$^{-12}$ s s$^{-1}$)        \dotfill &   4.8(2)                                                                 &   $-$0.7(6)                                                                &    $-$1.0(2)                                                                 &   $-$4.9(4)                                                                \\

\hline
\multicolumn{5}{c}{Derived Parameters}  \\
\hline
Angular offset from centre in $\alpha$, $\theta_{\alpha}$ (arcmin) \dotfill & +0.4179 & +0.0795 & +0.8326 & $-$0.1301 \\
Angular offset from centre in $\delta$, $\theta_{\delta}$ (arcmin) \dotfill & $-$0.4587 & +0.7636 & +0.4578 & +0.1712 \\
Total angular offset from centre, $\theta_{\perp}$ (arcmin) \dotfill & 0.6205 & 0.7677 & 0.9502 & 0.2150 \\
Total angular offset from centre, $\theta_{\perp}$ (core radii) \dotfill & 1.7882 & 2.2123 & 2.7383 & 0.6196 \\
Projected distance from centre, $r_{\perp}$ (pc) \dotfill & 0.8465 & 1.0473 & 1.2963 & 0.2933 \\
Spin Period, $P$ (ms)                                                  \dotfill &   3.53632915276243(3)                                    &   3.2103407093504(1)                                     &   4.0331811845726(2)                                     &   2.83040595787912(7)                                    \\
First Spin Period derivative, $\dot{P}$ ($10^{-21}$ s s$^{-1}$)                    \dotfill &   98.5103(5)   &   $-$1.830(1)  &   34.0076(6)  &   $-$120.541(1)    \\
Line of sight acceleration from cluster field, $a_{\ell, \rm GC}$ ($10^{-9}\, \rm m \, s^{-2}$)\dotfill & +7.31(32) & $-$1.0(0.9) & +3.0(0.7) & $-$14.2(1.1) \\
Intrinsic spin period derivative, $\dot{P}_{\rm int}$ (10$^{-21}$ s s$^{-1}$)      \dotfill &   11.9(3.7)                                                                  &   9(9)                                                                 &   $-$6(9)                                                                  &   13(10)                                                                  \\
Characteristic Age, $\tau_{\rm c}$ (Gyr)                              \dotfill &   4.7  &  $>$1.9  &  $>\,$5.0   & $>$1.3 \\
Line-of-sight jerk, $\dot{a}_\ell$ ($10^{-21}\, \rm m\, s^{-3}$) \dotfill & $-$3.10(26) & $-$154.5(2.2) & $-$0.9(1.3) & 6.6(6) \\
Mass Function, $f(M_{\rm p})$ (${\rm M}_\odot$) \dotfill & 0.0016409130(15) & 0.001927197(6) & 0.002374007(5) & 0.0003345154(10) \\
Minimum companion mass, $M_{\rm c, min}$ (${\rm M}_\odot$)            \dotfill &   0.159                                                               &   0.168                                                               &   0.181                                                               &   0.091                                                               \\
Median companion mass, $M_{\rm c, med}$ (${\rm M}_\odot$)             \dotfill &   0.185                                                               &   0.196                                                               &   0.212                                                               &   0.105                                                               \\
Total Mass, $M$ (M$_\odot$)                                           \dotfill &   2.3(7)                                                                  &   1.665(7)                                                                  &   --                                                                     &   3.1(1.1)                                                                  \\
\hline
\end{tabular} }
\end{center} 
\end{table*}

%%%%%%%%%%%%%%%%%%%%%%%%%%%%%%%%%%%%%%%%%%%%%%%%%%%%%%%%%%%%%%%
% TABLE OF WD-pulsar BINARIES - continued
%%%%%%%%%%%%%%%%%%%%%%%%%%%%%%%%%%%%%%%%%%%%%%%%%%%%%%%%%%%%%%%
\begin{table*}
\contcaption{}
\begin{center}{\scriptsize
\setlength{\tabcolsep}{6pt}
\renewcommand{\arraystretch}{1.3}
\begin{tabular}{l c c c}
\hline
Pulsar  & 47 Tuc T & 47 Tuc U & 47 Tuc Y  \\
\hline\hline
Right Ascension, $\alpha$ (J2000)                                     \dotfill &   00:24:08.5491(5)                                                       &   00:24:09.8366(1)                                                       &   00:24:01.4026(1)                                                       \\
Declination, $\delta$ (J2000)                                         \dotfill &   $-$72:04:38.932(3)                                                       &   $-$72:03:59.6882(4)                                                      &   $-$72:04:41.8363(4)                                                      \\
Proper Motion in $\alpha$, $\mu_\alpha$ (mas yr$^{-1}$)               \dotfill &   5.1(6)                                                                 &   4.6(2)                                                                 &   4.4(1)                                                                 \\
Proper Motion in $\delta$, $\mu_\delta$ (mas yr$^{-1}$)               \dotfill &   $-$2.6(7)                                                                &   $-$3.8(1)                                                                &   $-$3.4(1)                                                                \\
% Parallax (mas)                                                        \dotfill &   0.2132                                                                 &   0.2132                                                                 &   0.2132                                                                 \\
Spin Frequency, $f$ (Hz)                                        \dotfill &   131.77869947406(2)                                                     &   230.264772211776(6)                                                    &   455.23717843241(1)                                                     \\
First Spin Frequency derivative, $\dot{f}$ ($10^{-15}$ Hz s$^{-1}$) \dotfill &   $-$5.1021(2) &   $-$5.04916(9) &   7.2891(2) \\
Second Spin Frequency derivative, $\ddot{f}$ ($10^{-27}$ Hz s$^{-2}$)               \dotfill &   $-$3(2) &   18.8(6)   &   $-$21.1(9) \\
%Reference Epoch (MJD)                                                 \dotfill &   51600.000000                                                           &   51600.000000                                                           &   51600.000000                                                           \\
Start of Timing Data (MJD)                                            \dotfill &   50683.712                                                              &   48515.506                                                              &   50739.663                                                              \\
End of Timing Data (MJD)                                              \dotfill &   56466.934                                                              &   56466.919                                                              &   56508.973                                                              \\
Dispersion Measure, DM (pc cm$^{-3}$)                                 \dotfill &   24.41(2)                                                               &   24.337(4)                                                              &   24.468(4)                                                              \\
%Solar System Ephemeris                                                \dotfill &   DE421                                                                  &   DE421                                                                  &   DE421                                                                  \\
%Terrestrial Time Standard                                             \dotfill &   UTC(NIST)                                                              &   UTC(NIST)                                                              &   UTC(NIST)                                                              \\
%Time Units                                                            \dotfill &   TDB                                                                    &   TDB                                                                    &   TDB                                                                    \\
Number of ToAs                                                        \dotfill &   554                                                                    &   1309                                                                   &   804                                                                    \\
Residuals RMS ($\mu$s)                                                \dotfill &   54.36                                                                  &   9.68                                                                   &   8.11                                                                   \\

\hline
\multicolumn{4}{c}{Binary Parameters}  \\
\hline
Binary Model                                                          \dotfill &   ELL1                                                                   &   ELL1                                                                   &   ELL1                                                                   \\
Projected Semi-major Axis, $x_p$ (lt-s)                               \dotfill &   1.338501(5)                                                            &   0.5269494(7)                                                           &   0.6685965(7)                                                           \\
First Laplace-Lagrange parameter, $\eta$                              \dotfill &   3.55(7)$\times 10^{-4}$                                                &   $-$2.9(4)$\times 10^{-5}$                                                &   $-$3(3)$\times 10^{-6}$                                                  \\
Second Laplace-Lagrange parameter, $\kappa$                            \dotfill &   1.85(7)$\times 10^{-4}$                                                &   1.43(2)$\times 10^{-4}$                                                &   0(2)$\times 10^{-6}$                                                \\
Epoch of passage at Ascending Node, $T_\textrm{asc}$ (MJD)            \dotfill &   51600.5692696(7)                                                       &   51600.3893516(1)                                                       &   51554.8340067(2)                                                       \\
Rate of periastron advance, $\dot{\omega}$ (deg/yr)                   \dotfill &   --                                                                     &   1.17(32)                                                                 &   --                                                                     \\
Orbital Period, $P_{\rm b}$ (days)                                          \dotfill &   1.126176771(1)                                                         &   0.42910568324(8)                                                       &   0.5219386107(1)                                                        \\
Orbital Period derivative, $\dot{P}_b$ (10$^{-12}$ s s$^{-1}$)        \dotfill &   2.5(1.1)                                                                   &   0.66(5)                                                                &   $-$0.82(7)                                                               \\
\hline
\multicolumn{4}{c}{Derived Parameters}  \\
\hline
Angular offset from centre in $\alpha$, $\theta_{\alpha}$ (arcmin) \dotfill & +0.2215 & +0.3207 & $-$0.3283 \\
Angular offset from centre in $\delta$, $\theta_{\delta}$ (arcmin) \dotfill & +0.2280 & +0.8821 & +0.1799 \\
Total angular offset from centre, $\theta_{\perp}$ (arcmin) \dotfill & 0.3179 & 0.9386 & 0.3743 \\
Total angular offset from centre, $\theta_{\perp}$ (core radii) \dotfill & 0.9160 & 2.7049 & 1.0788 \\
Projected distance from centre, $r_{\perp}$ (pc) \dotfill & 0.4336 & 1.2805 & 0.5107 \\
Spin Period, $P$ (ms)                                                  \dotfill &   7.5884798073671(9)                                    &   4.3428266963923(1)                                     &   2.19665714352124(6)                                    \\
First Spin Period derivative, $\dot{P}$ ($10^{-21}$ s s$^{-1}$)                    \dotfill &   293.80(1)  &   95.228(2)                                             &   $-$35.1720(8)  \\
Line of sight acceleration from cluster field, $a_{\ell, \rm GC}$ ($10^{-9}\, \rm m \, s^{-2}$) \dotfill & 7.7(3.5) & 5.31(38) & $-$5.4(4) \\
Intrinsic spin period derivative, $\dot{P}_{\rm int}$ (10$^{-21}$ s s$^{-1}$)      \dotfill &  99(89) & 18(5) &  4.7(3.3) \\
Characteristic Age, $\tau_{\rm c}$ (Gyr)                              \dotfill &  $>$0.43 &  3.8  & $>$3.1 \\
Line-of-sight jerk, $\dot{a}_\ell$ ($10^{-21}\, \rm m\, s^{-3}$) \dotfill & 7.8(3.7) & $-$24.4(8) & 13.9(6) \\
Mass Function, $f(M_{\rm p})$ (${\rm M}_\odot$)                       \dotfill &   0.002030139(25)                                                               &     0.0008532200(35)             &   0.0011779754(37)                                                               \\
Minimum companion mass, $M_{\rm c, min}$ (${\rm M}_\odot$)            \dotfill &   0.171                                                               &   0.126                                                               &   0.141                                                               \\
Median companion mass, $M_{\rm c, med}$ (${\rm M}_\odot$)             \dotfill &   0.200                                                               &   0.146                                                               &   0.164                                                               \\
Total Mass, $M$ (M$_\odot$)                                           \dotfill &   --                                                                     &   1.7(7)                                                                  &   --                                                                     \\
\hline
\end{tabular} }
\end{center} 
\end{table*}

%%%%%%%%%%%%%%%%%%%%%%%%%%%%%%%%%%%%%%%%%%%%%%%%%%%%%%%%%%%%%%%
% TABLE OF BLACK WIDOW/REDBACK PULSARS
%%%%%%%%%%%%%%%%%%%%%%%%%%%%%%%%%%%%%%%%%%%%%%%%%%%%%%%%%%%%%%%

\begin{table*}
\caption{Timing parameters for the four black-widow systems with timing solutions in 47 Tuc, as obtained from fitting the observed ToAs with {\sc TEMPO} (for the fifth black widow system, 47~Tuc~P, it was possible to derive a precise orbit but no phase-coherent solution, see Paper I). For the characteristic age, we either present the median or a 2-$\sigma$ lower limit. The orbital models used are the ELL1 (Lange et al. 2001) and BTX (D. Nice, unpublished); for the latter the orbital periods are derived from the orbital frequency and presented in square brackets.}
\label{tab:MSP_BW}
\begin{center}{\scriptsize
\setlength{\tabcolsep}{6pt}
\renewcommand{\arraystretch}{1.3}
\begin{tabular}{l c c c c}
\hline
Pulsar & 47 Tuc I & 47 Tuc J & 47 Tuc O & 47 Tuc R \\
\hline\hline
Right Ascension, $\alpha$ (J2000)                                     \dotfill &   00:24:07.9347(2)                                                       &   00:23:59.4077(1)                                                       &   00:24:04.65254(6)                                                      &   00:24:07.5851(2)                                                       \\
Declination, $\delta$ (J2000)                                         \dotfill &   $-$72:04:39.6815(7)                                                      &   $-$72:03:58.7908(5)                                                      &   $-$72:04:53.7670(2)                                                      &   $-$72:04:50.3954(5)                                                      \\
Proper Motion in $\alpha$, $\mu_\alpha$ (mas yr$^{-1}$)               \dotfill &   5.0(2)                                                                 &   5.27(6)                                                                &   5.01(5)                                                                &   4.8(1)                                                                 \\
Proper Motion in $\delta$, $\mu_\delta$ (mas yr$^{-1}$)               \dotfill &   $-$2.1(2)                                                                &   $-$3.59(9)                                                               &   $-$2.58(8)                                                               &   $-$3.3(2)                                                                \\
%Parallax (mas)                                                        \dotfill &   0.2132                                                                 &   0.2132                                                                 &   0.2132                                                                 &   0.2132                                                                 \\
Spin Frequency, $f$ (Hz)                                        \dotfill &   286.94469953049(1)                                                     &   476.04685844061(1)                                                     &   378.308788360098(6)                                                    &   287.31811946930(1)                                                     \\
First Spin Frequency derivative, $\dot{f}$ ($10^{-15}$ Hz s$^{-1}$) \dotfill &   3.7771(2) &   2.2190(2) & $-$4.34352(8) &  $-$12.2467(2) \\
Second Spin Frequency derivative, $\ddot{f}$ ($10^{-27}$ Hz s$^{-2}$)               \dotfill &   $-$33.5(9) &   20(1) &   43.8(5) &   $-$8.5(1.5) \\
%Reference Epoch (MJD)                                                 \dotfill &   51600.000000                                                           &   51600.000000                                                           &   51600.000000                                                           &   51600.000000                                                           \\
Start of Timing Data (MJD)                                            \dotfill &   47859.462                                                              &   47717.894                                                              &   50683.712                                                              &   50742.607                                                              \\
End of Timing Data (MJD)                                              \dotfill &   56466.878                                                              &   56388.106                                                              &   56508.991                                                              &   55362.896                                                              \\
Dispersion Measure, DM (pc cm$^{-3}$)                                 \dotfill &   24.43(1)                                                               &   24.588(3)                                                              &   24.356(2)                                                              &   24.361(7)                                                              \\
%Solar System Ephemeris                                                \dotfill &   DE421                                                                  &   DE421                                                                  &   DE421                                                                  &   DE421                                                                  \\
%Terrestrial Time Standard                                             \dotfill &   UTC(NIST)                                                              &   UTC(NIST)                                                              &   UTC(NIST)                                                              &   UTC(NIST)                                                              \\
%Time Units                                                            \dotfill &   TDB                                                                    &   TDB                                                                    &   TDB                                                                    &   TDB                                                                    \\
Number of ToAs                                                        \dotfill &   1201                                                                   &   10135                                                                  &   1903                                                                   &   449                                                                    \\
Residuals RMS ($\mu$s)                                                \dotfill &   18.26                                                                  &   4.89                                                                   &   9.70                                                                   &   10.81                                                                  \\

\hline
\multicolumn{5}{c}{Binary Parameters}  \\
\hline
Binary Model                                                          \dotfill &   ELL1                                                                   &   BTX                                                                    &   BTX                                                                    &   ELL1                                                                   \\
Projected Semi-major Axis, $x_p$ (lt-s)                               \dotfill &   3.8446(1)$\times 10^{-2}$                                              &   4.04058(6)$\times 10^{-2}$                                             &   4.51533(3)$\times 10^{-2}$                                             &   3.3363(1)$\times 10^{-2}$                                              \\
Orbital Eccentricity, $e$                                             \dotfill &   --                                                                     &   0                                                           &   0                                                           &   --                                                                     \\
Longitude of Periastron, $\omega$ (deg)                               \dotfill &   --                                                                     &   0                                                         &   0                                                         &   --                                                                     \\
Epoch of passage at Periastron, $T_0$ (MJD)                           \dotfill &   --                                                                     &   51600.1084250(6)                                                       &   51600.0757563(3)                                                       &   --                                                                     \\
First Laplace-Lagrange parameter, $\eta$                              \dotfill &   0                                                           &   --                                                                     &   --                                                                     &   $-$10(6)$\times 10^{-5}$                                                 \\
Second Laplace-Lagrange parameter, $\kappa$                            \dotfill &   0                                                           &   --                                                                     &   --                                                                     &   $-$3(7)$\times 10^{-5}$                                                  \\
Epoch of passage at Ascending Node, $T_\textrm{asc}$ (MJD)            \dotfill &   51600.002421(2)                                                        &   --                                                                     &   --                                                                     &   51600.0029871(6)                                                       \\
Orbital Period, $P_{\rm b}$ (days)                                          \dotfill &   0.2297922489(4)                                                        &   [ 0.12066493766(13) ]                                                       &
[ 0.13597430589(9) ] &   6.623147751(6)$\times 10^{-2}$                                         \\
Orbital Period derivative, $\dot{P}_b$ (10$^{-12}$ s s$^{-1}$)        \dotfill &   $-$0.8(2)                                                                &   --                                                                     &   --                                                                     &   0.19(4)                                                                \\
Orbital Frequency, $f_b$ (s$^{-1}$)                                   \dotfill &   --                                                                     &   9.59191153(1)$\times 10^{-5}$                                          &   8.511956725(6)$\times 10^{-5}$                                         &   --                                                                     \\
1st Orbital Freq. derivative, $f^{(1)}_b$ (s$^{-2}$)                  \dotfill &   --                                                                     &   $-$21(34)$\times 10^{-22}$                                                 &   $-$7.3(1)$\times 10^{-20}$                                               &   --                                                                     \\
2nd Orbital Freq. derivative, $f^{(2)}_b$ (s$^{-3}$)                  \dotfill &   --                                                                     &   $-$19(21)$\times 10^{-29}$                                                 &   $-$10(2)$\times 10^{-29}$                                                &   --                                                                     \\
3rd Orbital Freq. derivative, $f^{(3)}_b$ (s$^{-4}$)                  \dotfill &   --                                                                     &   3.9(5)$\times 10^{-35}$                                                &   33(15)$\times 10^{-37}$                                                &   --                                                                     \\
4th Orbital Freq. derivative, $f^{(4)}_b$ (s$^{-5}$)                  \dotfill &   --                                                                     &   $-$15(24)$\times 10^{-44}$                                                 &   --                                                                     &   --                                                                     \\
5th Orbital Freq. derivative, $f^{(5)}_b$ (s$^{-6}$)                  \dotfill &   --                                                                     &   $-$5.5(8)$\times 10^{-50}$                                               &   --                                                                     &   --                                                                     \\
6th Orbital Freq. derivative, $f^{(6)}_b$ (s$^{-7}$)                  \dotfill &   --                                                                     &   58(23)$\times 10^{-59}$                                                  &   --                                                                     &   --                                                                     \\
7th Orbital Freq. derivative, $f^{(7)}_b$ (s$^{-8}$)                  \dotfill &   --                                                                     &   55(10)$\times 10^{-66}$                                               &   --                                                                     &   --                                                                     \\
8th Orbital Freq. derivative, $f^{(8)}_b$ (s$^{-9}$)                  \dotfill &   --                                                                     &   $-$89(20)$\times 10^{-74}$                                                 &   --                                                                     &   --                                                                     \\
9th Orbital Freq. derivative, $f^{(9)}_b$ (s$^{-10}$)                 \dotfill &   --                                                                     &   $-$3.9(8)$\times 10^{-80}$                                               &   --                                                                     &   --                                                                     \\
10th Orbital Freq. derivative, $f^{(10)}_b$ (s$^{-11}$)               \dotfill &   --                                                                     &   86(15)$\times 10^{-89}$                                                  &   --                                                                     &   --                                                                     \\
11th Orbital Freq. derivative, $f^{(11)}_b$ (s$^{-12}$)               \dotfill &   --                                                                     &   15(37)$\times 10^{-96}$                                                &   --                                                                     &   --                                                                     \\
12th Orbital Freq. derivative, $f^{(12)}_b$ (s$^{-13}$)               \dotfill &   --                                                                     &   $-$42(8)$\times 10^{-104}$                                                    &   --                                                                     &   --                                                                     \\
\hline
\multicolumn{5}{c}{Derived Parameters}  \\
\hline
Angular offset from centre in $\alpha$, $\theta_{\alpha}$ (arcmin) \dotfill & +0.1742 & $-$0.4821 & $-$0.0783 & +0.1473 \\
Angular offset from centre in $\delta$, $\theta_{\delta}$ (arcmin) \dotfill & +0.2156 & +0.8972 & $-$0.0192 & +0.0371 \\
Total angular offset from centre, $\theta_{\perp}$ (arcmin) \dotfill & 0.2772 & 1.0185 & 0.0806 & 0.1519 \\
Total angular offset from centre, $\theta_{\perp}$ (core radii) \dotfill & 0.7989 & 2.9352 & 0.2322 & 0.4378 \\
Projected distance from centre, $r_{\perp}$ (pc) \dotfill & 0.3782 & 1.3895 & 0.1099 & 0.2072 \\
Spin Period, $P$ (ms)                                                  \dotfill &   3.4849920616629(1)                                     &   2.10063354535248(6)                                    &   2.64334329724356(4)                                    &   3.4804627074933(2)                                     \\
First Spin Period derivative, $\dot{P}$ ($10^{-21}$ s s$^{-1}$)                    \dotfill &   $-$45.873(2)  &   $-$9.7919(9)   &   30.3493(6)  &   148.351(3) \\
Line of sight acceleration from cluster field, $a_{\ell, \rm GC}$ ($10^{-9}\, \rm m \, s^{-2}$) \dotfill & $-$11.8(3.7) & -- & -- & 10.1(1.9) \\
Intrinsic spin period derivative, $\dot{P}_{\rm int}$ (10$^{-21}$ s s$^{-1}$)      \dotfill & 92(43) &   --   &   --  &   31(22)  \\
Characteristic Age, $\tau_{\rm c}$ (Gyr)                              \dotfill &   0.60    &   --   &   --   &  $>\,$0.73 \\
Line-of-sight jerk, $\dot{a}_\ell$ ($10^{-21}\, \rm m\, s^{-3}$) \dotfill & 35.0(1.0) & $-$12.5(9) & $-$34.68(37) & 8.9(1.5) \\
Mass Function, $f(M_{\rm p})$ (${\rm M}_\odot$)                       \dotfill &   0.0000011555(1)                                                               &     0.0000048646(2)                                                             & 
0.0000053461(1)                                                               &   0.0000090898(10)                                                               \\
Minimum companion mass, $M_{\rm c, min}$ (${\rm M}_\odot$)            \dotfill &   0.0132                                                               &   0.0214                                                               &   0.0221                                                               &   0.0264                                                               \\
Median companion mass, $M_{\rm c, med}$ (${\rm M}_\odot$)             \dotfill &   0.0153                                                               &   0.0248                                                               &   0.0256                                                               &   0.0306                                                               \\

\hline
\end{tabular} }
\end{center} 
\end{table*}

\begin{figure*}
	\includegraphics[width=0.6\textwidth]{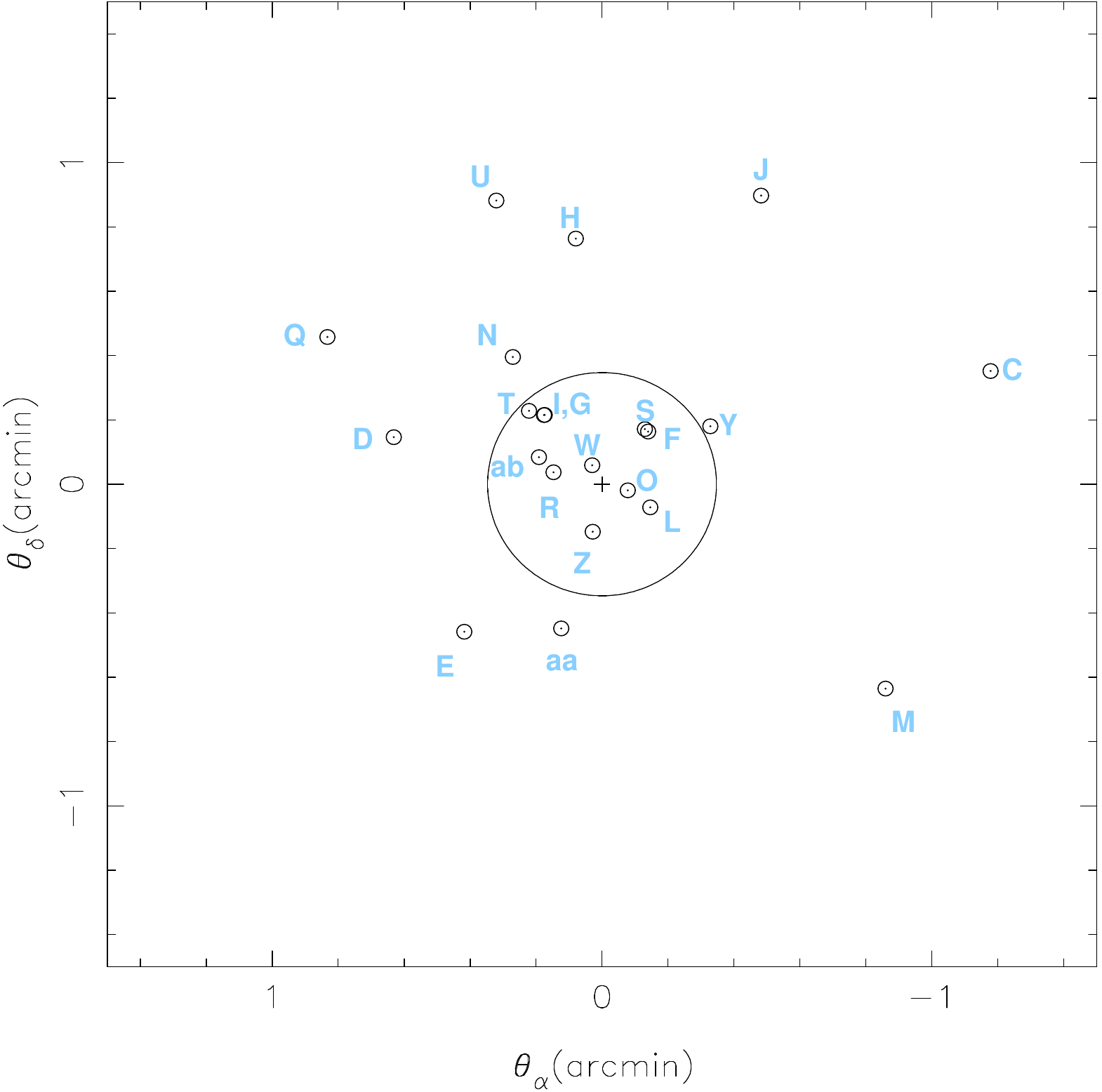}
    \caption{East-west ($\theta_{\alpha}$) and north-south ($\theta_{\delta}$)
    angular offsets from the centre of the GC 47~Tucanae for 22 of its 25 known pulsars.
    The central circle indicates the core radius. The 23$^{\rm rd}$ pulsar with a timing
    solution, 47~Tuc~X, is well outside the limits of this figure and its position relative to the
    cluster and the other pulsars is displayed graphically in Paper I \citep{rfc+16}.}
    \label{fig:positions}
\end{figure*}

\begin{figure*}
	\includegraphics[width=0.6\textwidth]{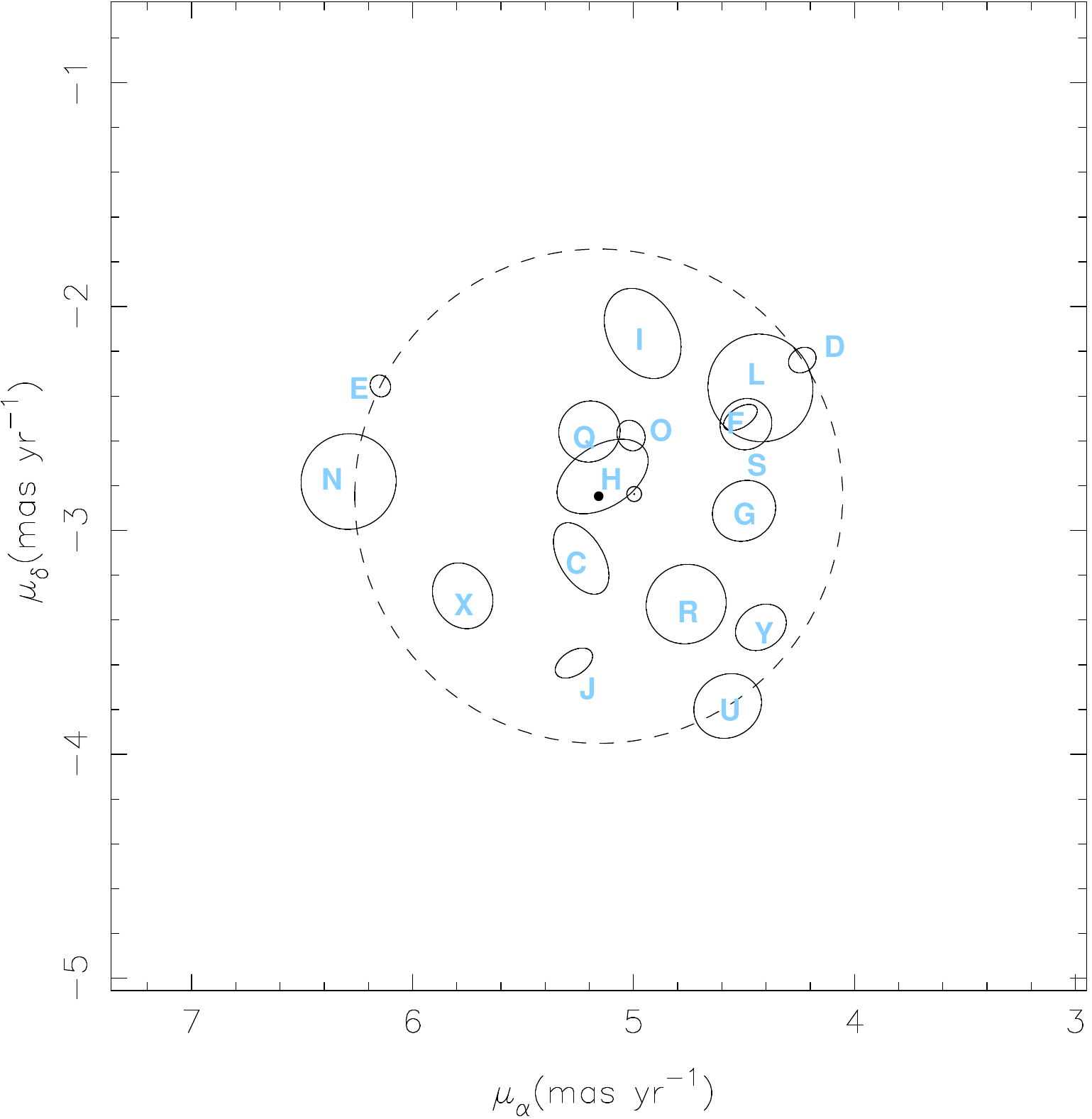}
    \caption{Proper motions for the 17 pulsars in 47~Tucanae where both 1-$\sigma$ uncertainties
    are smaller than 0.3 mas yr$^{-1}$. The proper motions are displayed by ellipses where the
    semi-axes have length of, and are aligned with, the 1-$\sigma$ proper motion
    uncertainties in ecliptic coordinates (where the positional
    and proper motion uncertainties are well defined). The differences in transverse velocity
    between the pulsars are highly significant, particularly
    for precisely timed pulsars like 47~Tuc~D, E, F, J and O.
    The average of all proper motions in $\alpha$ and $\delta$
    ($\mu_{\alpha}\, = \, 5.00 \, \rm  mas\, yr^{-1}$ and
     $\mu_{\delta}\, = \, -2.84 \, \rm  mas\, yr^{-1}$) is given by the Solar symbol
    near the centre of the plot. This represents an estimate of the motion of the GC as a whole.
    The dashed circle represents the minimal possible velocity
    envelope for these pulsars, its centre (the solid dot at the centre of the plot, at
    $\mu_{\alpha}\, = \, 5.16 \, \rm mas \, yr^{-1}$,
    $\mu_{\delta} \, = \, -2.85 \, \rm mas \, yr^{-1}$) represents another estimate of the proper
    motion of the globular cluster. This circle has a radius of 1.10 mas yr$^{-1}$.
    At the assumed distance to 47~Tuc (4.69 kpc) this is a velocity relative to the GC of
    24.5 km s$^{-1}$, or about half the escape velocity from the centre of the cluster.}
    \label{fig:pms}
\end{figure*}

\begin{figure*}
	\includegraphics[width=0.6\textwidth]{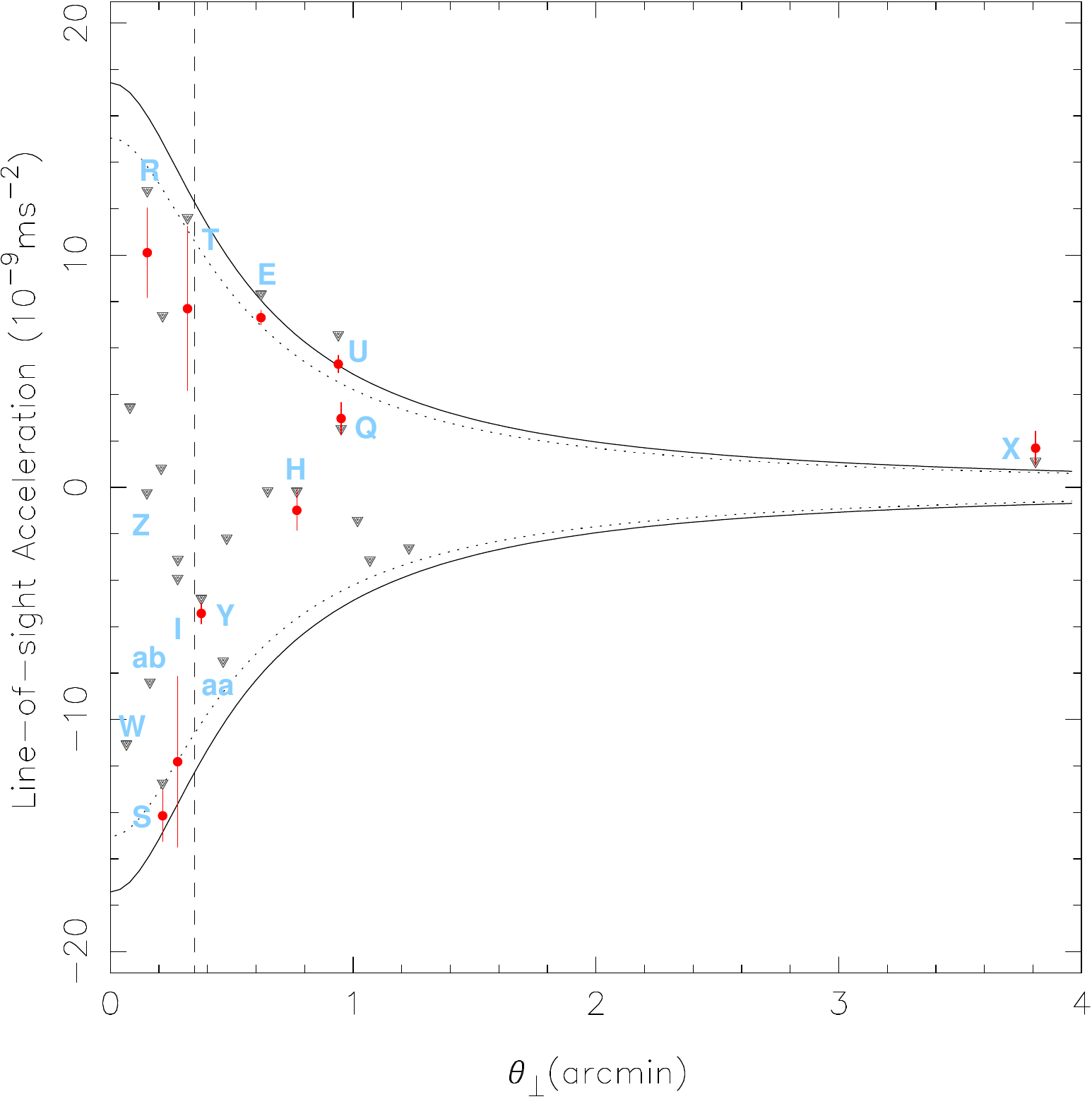}
    \caption{Line of sight accelerations ($a_\ell$) as a function of the total angular offset from the centre 
    of the cluster ($\theta_{\perp}$) for the pulsars in 47~Tuc. The inverted triangles represent, for
    each pulsar system, an upper limit for its acceleration
    in the field of the cluster, this is determined from $\dot{P}_{\rm obs}$
    (see discussion in Section~\ref{sec:accelerations}).
    This is not a measurement of the real acceleration in the field of the cluster because of
    a contribution from the intrinsic spin period derivative of each pulsar ($\dot{P}_{\rm int}$).
    The red error bars represent measurements of the line-of-sight accelerations of 10 binary pulsars
    (47~Tuc~E, H, I, Q, R, S, T, U, X and Y, which are named) determined from their orbital period
    derivatives, $\dot{P}_{\rm b, obs}$.  We also plot the maximum and minimum
    accelerations ($a_{\ell,\, \rm GC, max}$) along each line of sight predicted by the
    analytical model of the cluster described in Section~\ref{sec:jerks}, with distances of
    4.69 kpc (solid lines) and 4.15 kpc (dotted lines). We also name the systems with
    recently determined timing solutions (R, W, Z, aa, ab).
    The core radius is indicated by the vertical dashed line.}
\label{fig:accel}
\end{figure*}

\begin{figure*}
	\includegraphics[width=0.6\textwidth]{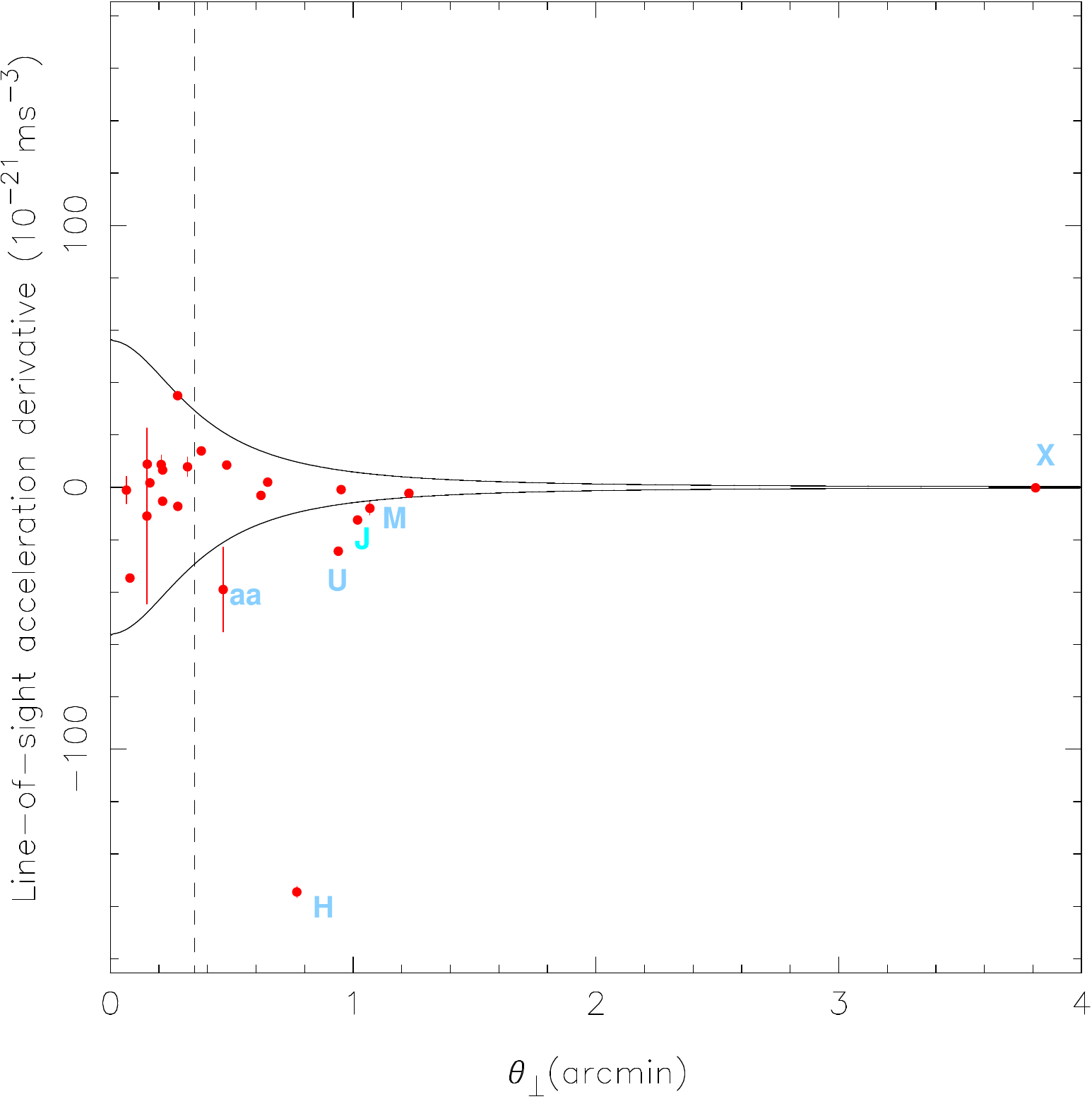}
    \caption{The red dots and error bars represent
    measurements and uncertainties of the line-of-sight jerks ($\dot{a}_\ell$)
    for all the pulsars in 47~Tuc. The solid lines display, for each line of sight
    $\theta_{\perp}$, the maximum and minimum theoretical
    expectations (from Eq.~\ref{eq:a-dot}) for the line-of-sight jerks
    caused by the motion of pulsars in the mean field of the cluster
    ($\dot{a}_{\ell,\, \rm GC, max}$).
    For some pulsars (47~Tuc~H and U and possibly 47~Tuc~J),
    the observed $\dot{a}_\ell$ is larger than $\dot{a}_{\ell,\, \rm GC, max}$;
    this is likely due to the presence of stars near those systems. The core radius
    is indicated by the vertical dashed line.}
    \label{fig:accel-dot}
\end{figure*}

\section{Proper motions}
\label{sec:proper_motions}

In Fig.~\ref{fig:positions} we display the angular offsets of 22 pulsars in the sky relative to
the centre of the cluster. Their numerical values are presented
in Tables~\ref{tab:isolated} to \ref{tab:MSP_BW}.
The figure does not display 47~Tuc~X, which is at a distance of $\sim \, 3.8 \arcmin$ from the centre of the cluster (Paper I).
All other pulsars (including those with new timing solutions) are well 
within the distance of 47~Tuc~C to the centre of the cluster, 1.22\arcmin.
The Parkes 20 cm beam has a half-power radius of 7\arcmin, so this is not a
selection effect. As shown by \cite{hge+05}, the real cause is
{\em mass segregation}: this close to the centre the relaxation time
is much shorter than the age of the cluster, hence, the pulsars are likely to have
reached dynamical equilibrium with the stellar population there.

One of the main benefits of long-term timing is a better determination of the proper motions.
In \cite{fcl+01}, the number and precision of proper motions was small and only the
motion of the GC as a whole was detectable. With a few more years of intense timing
with the hAFB, some of the proper motions were measured precisely enough to detect
relative motions, particularly for 47~Tuc~D, E  and J \citep{fck+03}.

Because of the increased timing baselines, the proper motions
presented in this work (depicted graphically in Fig.~\ref{fig:pms}) are significantly more precise. Although the proper motions themselves are displayed
in J2000 equatorial coordinates, the error ellipses are aligned according to ecliptic
coordinates, where the measurement uncertainties are least correlated. In Fig.~\ref{fig:pms},
we display the 17 pulsars for which both proper motion 1-$\sigma$ Monte Carlo
uncertainties (in ecliptic longitude, $\lambda$, and ecliptic latitude, $\beta$) are
smaller than $0.3 \, \rm mas \, yr^{-1}$.
For these pulsars, the (unweighted) average proper motion is 
$\mu_{\alpha}\, = \, 5.00 \, \rm  mas\, yr^{-1}$ and
$\mu_{\delta}\, = \, -2.84 \, \rm  mas\, yr^{-1}$, and is 
depicted by the Solar symbol in Fig.~\ref{fig:pms}.
This represents the simplest estimate for the proper motion of the cluster as
a whole, and it is consistent with the estimate presented in
\cite{fck+03}: $\mu_{\alpha}\, = \, 5.3 \, \pm \, 0.6 \, \rm  mas\, yr^{-1}$ and
$\mu_{\delta}\, = \, -3.3 \, \pm \, 0.6\, \rm  mas\, yr^{-1}$.
The standard deviations of the proper motions around this average ($\sigma_{\mu}$) are
$0.59 \, \rm  mas\, yr^{-1}$ in $\alpha$ and $0.49 \, \rm  mas\, yr^{-1}$ in $\delta$.
At a distance of 4.69 kpc \citep{wgk+12}, these standard deviations correspond
to 13.2 and $10.9\, \rm km\, s^{-1}$ respectively. The uncertainty in the mean value is
given by $\sigma_{\mu} / \sqrt{N}$, where $N$ is the number of measurements (17, in this case).
Thus our uncertainties for the mean cluster motion are $\sigma_{\mu_{\alpha}}\, = \, 0.14\, \rm mas \, yr^{-1}$
and  $\sigma_{\mu_{\delta}}\, = \, 0.12\, \rm mas \, yr^{-1}$.

An alternative method to estimate the overall motion of the cluster is to
require that all observed pulsar proper motions fit within the smallest possible
velocity envelope.
This corresponds to finding the centre of a circle defined by the
proper motions of the three outermost pulsars in the proper motion plot, namely 
47~Tuc~D, E and U (alternatively we can use 47~Tuc~D, N and U; we prefer the
former set because the proper motion for 47~Tuc~E is known much more precisely).
This minimal proper motion envelope is represented by the dashed circle and has a
radius of $1.10 \, \rm  mas\, yr^{-1}$; at 4.69 kpc this represents a
velocity of $24.5\, \rm km\, s^{-1}$, or about half of the
escape velocity from the centre of the cluster
($\sim \, 50\, \rm km\, s^{-1}$, e.g., \citealt{mam+06}).
The centre of this minimal envelope is at
$\mu_{\alpha}\, = \, 5.16 \, \rm  mas\, yr^{-1}$ and
$\mu_{\delta}\, = \, -2.85 \, \rm  mas\, yr^{-1}$,  represented by a solid 
dot at the centre of Fig.~\ref{fig:pms}. This $\mu_{\alpha}$ is almost 1-$\sigma$
consistent with the average estimated above, the $\mu_{\delta}$ is
practically identical to the average.

\subsection{Comparison with optical proper motions}

We now compare these numbers with previous literature. 
Regarding the absolute proper motion, the latest relevant study combines {\em HIPPARCOS} and
{\em GAIA} positions to derive absolute proper motions for five Galactic
globular clusters, among which is 47~Tuc \citep{wm16}. The values they
obtain ($\mu_{\alpha}\, = \, 5.50 \, \pm \, 0.70\, \rm  mas\, yr^{-1}$ and
$\mu_{\delta}\, = \, -3.99 \, \pm \, 0.55 \, \rm  mas\, yr^{-1}$)
are consistent with  our measurement of the average proper motion in $\alpha$,
but in $\delta$ the deviation is $-$2.1 $\sigma$, i.e., only marginally consistent.
In Section 3.1 of that paper they list previous measurements of the
proper motion of 47~Tuc and discuss their consistency, and it is clear that
there is some disagreement between the proper motion estimates obtained by 
different methods. The situation will likely improve significantly with the
second release of {\em GAIA} data.

Our 1-D standard deviations for the proper motions agree with the
$\sigma_{\mu, 0}$ obtained by \cite{wmba15}.
This result agrees qualitatively with the observation
by \cite{mam+06} that the observed velocity
dispersion is largely constant across magnitude
range, i.e., it appears to be the same for stellar populations of different
masses. 

\subsection{Proper motion pairs?}

Given the extreme proximity of 47~Tuc~I and G in the sky and
in acceleration, there is a suggestion that these pulsars could be in a bound
system, with a semi-major axis $a_{\rm p}$ of at least 600 a.u. \citep{fcl+01}.
Such systems are not stable in 47~Tuc, since their cross section for violent interactions 
is too large, but they can exist temporarily.
If this were the case, then the maximum relative orbital velocity should be of the order of
$v\, \sim \, \sqrt{G M / a_{\rm p}}\, = \, 2 \, \rm km \, s^{-1}$. At the
distance of 47~Tuc, this translates to an upper limit on the
difference of proper motions of about $0.09 \, \rm mas\, yr^{-1}$.
As we can see from Fig.~\ref{fig:pms}, the difference is of the order
of $1 \, \rm mas\, yr^{-1}$, ten times larger. We conclude therefore that,
despite their proximity, these two pulsars are not in a bound system.

Two other pulsars, 47~Tuc~F and S, are also remarkably close to
each other and have identical DMs. In this case the minimum separation
is 3700 a.u., requiring a maximum relative orbital velocity
$v\, \sim \, 0.8 \, \rm km \, s^{-1}$ and a maximal proper
motion difference of $0.036 \, \rm mas\, yr^{-1}$. 
Interestingly, this is not excluded by our measurements:
as we can see in Fig.~\ref{fig:pms}, the proper motion
of 47~Tuc~F falls within the 1-$\sigma$ contour for
the proper motion of 47~Tuc~S. The latter covers
1.1\% of the proper motion surface within the velocity envelope determined
above, so that is the probability of coincidence for any given
pulsar. Given their spatial proximity, the proper motion coincidence
is suggestive of a temporarily bound status.

As mentioned in \cite{fcl+01}, another test of the bound nature of these
systems would be the detection of changes in their line-of-sight accelerations,
which will produce a second derivative of the spin frequency, $\ddot{f}$.
However, as we shall see in Section~\ref{sec:jerks}, the $\ddot{f}$'s of these
pulsars can be accounted for by their movement in the cluster potential.

\section{Spin period derivatives}
\label{sec:jerks}

We will now discuss the measurements of the spin frequency derivatives of the pulsars
in 47~Tucanae. Like those of other pulsars in GCs
(and unlike the spin period derivatives observed in the Galactic disk)
the first spin frequency derivatives for the pulsars in 47~Tuc are mostly caused by
dynamical effects.
Higher spin frequency derivatives are, within our timing precision, caused
entirely by dynamical effects.

\subsection{First spin period derivative and upper limits on acceleration in the cluster field}
\label{sec:pdots}

The observed variation of spin period
$\dot{P}_{\rm obs}$ is generally given by the following equation:
\begin{equation}
\frac{\dot{P}_{\rm obs}}{P} \, = \, \frac{\dot{P}_{\rm int}}{P} \, + \, \frac{\mu^2 d}{c} + \frac{a_{\ell,\, \rm GC}}{c} + \frac{a}{c},
\label{eq:p_dot}
\end{equation}
where $P$ is the observed pulsar spin period, $\dot{P}_{\rm obs}$ is the observed spin period derivative,
$\dot{P}_{\rm int}$ is the intrinsic spin period derivative,
$\mu$ is the composite proper motion, $d$ is the distance to the cluster
(the term $\mu^2 d/c$ is known as the Shklovskii effect, see \citealt{shk70}),
$c$ is the speed of light, $a_{\ell, \, \rm GC}$ is the line-of-sight acceleration
of the pulsar in the gravitational field of the cluster and $a$ is the
difference between the accelerations of the Solar System and 47 Tuc in the 
field of the Galaxy, projected along the direction to 47~Tuc
($a \, = \, -1.172 \, \times \, 10^{-10}\, \rm m\, s^{-2}$, calculated using
the \citealt{rmb+14} model for the Galactic rotation). In principle this
equation could have other contributions, in particular accelerations
caused by nearby stars. However, as shown by \cite{phi93}, even in dense
clusters those are very rarely relevant. As we shall see, the
dominant term for the pulsars in 47~Tuc is $a_{\ell, \, \rm GC}$.

For most pulsars, we can only derive an upper limit on this dominant term
from $\dot{P}_{\rm obs}/P$, since $\dot{P}_{\rm int}$ is generally not known but is always positive:
\begin{equation}
a_{\ell, \rm max} \, \doteq \, a_{\ell, \, \rm GC} + \frac{\dot{P}_{\rm int}}{P}c \, = \, \frac{\dot{P}_{\rm obs}}{P} c - \mu^2 d - a,
\label{eq:almax}
\end{equation}
these are displayed graphically in Fig.~\ref{fig:accel} as the triangles pointing down (to emphasise that
they represent an upper limit on the cluster acceleration).

The solid lines represent the maximum line-of-sight acceleration due to the cluster potential
($a_{\ell, \, \rm GC, \, \rm max}$) for the analytical model of the cluster
described in \cite{fhn+05}.
This uses the mass distribution presented in \cite{king62} for the case where
we are near (within $\sim 4$ core radii of) the centre of the cluster:
\begin{equation}
\label{eq:density}
\rho(x) \, = \, \frac{\rho(0)}{(1 + x^2)^{3/2}},
\end{equation}
where $x$ is the distance to the centre divided by the core radius $r_c = \theta_c d$.
In this paper we use the $\rho(0)$ defined in eq.~\ref{eq:central_density}.
Since this is independent of distance, the
mass of the cluster within a particular angular distance (i.e., within $x$ core radii,
$M_{\rm GC}(x)$) is proportional to $d^3$, so the acceleration
at that $x$ is proportional to $M_{\rm GC}(x) / d^2$, i.e., proportional to $d$:
\begin{equation}
\label{eq:accel2}
a_{\rm GC}(x) = \frac{9 \sigma^2_{\mu, 0}\, d}{\theta_c} \frac{1}{x^2}
\left( \frac{x}{\sqrt{1+x^2}}  - \sinh^{-1}x \right).
\end{equation}
The line-of-sight component of this acceleration, $a_{\ell,\, \rm GC}(x)$, can
be obtained by multiplying $a_{\rm GC}(x)$ by $\ell /x$, where $\ell$ is the
distance (also in core radii) to the plane of the sky that passes through the
centre of the cluster ($\Pi$), such that
$x\, = \, \sqrt{\ell ^2 \, +\, x_{\perp}^2}$
and $x_{\perp}\, = \, r_{\perp} / r_c \, \equiv \, \theta_{\perp} / \theta_c$.
For each pulsar line-of-sight (characterized by a constant angular offset from
the centre, $\theta_{\perp}$), we calculate
$a_{\ell, \, \rm GC}(x)$ for a variety of values of $\ell$,
recording the maximum and minimum values found, $a_{\ell, \, \rm GC, max}$;
these are the lines displayed in Fig.~\ref{fig:accel}.
For the line of sight going through the centre, we obtain the largest possible
acceleration induced by the field of the cluster:
\begin{equation}
a_{\ell, \rm GC, max}(0) = 1.5689 \frac{\sigma^2_{\mu, 0}\, d}{\theta_c};
\end{equation}
the numerical factor matches the more general expectation of
$1.50 \pm 0.15$ from Eq.~3.6 in \cite{phi93}. The latter was used
to constrain the cluster parameters in \cite{fck+03}.

Apart from $d$, the predicted $a_{\ell, \rm GC, max}(x_{\perp})$
depend only on unambiguous angular measurements,
this means that measurements of pulsar accelerations can be used to constrain $d$,
i.e., this is a second kinematic distance measurement.

None of the pulsars, including those with recently published solutions
(47~Tuc~R, W, X, Y, Z, aa and ab, all named in Fig.~\ref{fig:accel})
has a value of $a_{\ell, \rm max}$ that is significantly
larger than the model $a_{\ell,\, \rm GC, \, \rm max}$ for its line of sight.
The magnitude of the $a_{\ell,\, \rm GC}$ must be
significantly larger than the contributions from $\dot{P}_{\rm int}$,
otherwise a majority of $\dot{P}_{\rm obs}$
would be positive, while in fact similar numbers of pulsars have negative and
positive $\dot{P}_{\rm obs}$. For three pulsars,
47~Tuc~E, U and X, the $a_{\ell, \rm max}$ is slightly larger
than $a_{\ell,\, \rm GC, \, \rm max}$. For 47~Tuc~E and U, this is caused by the contribution
of their $\dot{P}_{\rm int}$; as we will see in Section~\ref{sec:accelerations},
their line-of-sight accelerations are (just about)
consistent with the cluster model. For 47~Tuc~X, it is likely that the
same is happening, although in that case the $\dot{P}_{\rm b,\, obs}$ is not yet precise
enough to reach any definite conclusions.
However, it is unlikely that the analytical acceleration model described above
is still entirely valid at its large $\theta_{\rm \perp}$.

\subsection{Second spin frequency derivative and jerk}

For the vast majority of MSPs observed in the Galactic disk
there is no detectable timing noise, even with timing precision much
better than what we can achieve for the MSPs in 47~Tuc. This means that
the large second spin frequency derivatives ($\ddot{f}$) observed
for the latter are much more likely to reflect their rate of change of
${a}_\ell$, normally known as the (line-of-sight) ``jerk'' ($\dot{a}_\ell$).
Rearranging equation (2) in \citep{jr97}, we get:
\begin{equation}
\frac{\dot{a}_\ell}{c} \, = \, \left( \frac{\dot{f}}{f} \right)^2 \, - \, \frac{\ddot{f}}{f}\, \simeq \,  -\ddot{f}P,
\label{eq:F2}
\end{equation}
the  approximation is valid since the first term, $(\dot{f}/ f)^2$,
is many orders of magnitude smaller than $\ddot{f}/f$.
According to \cite{phi92}, $\dot{a}_\ell$ has two main physical contributions.
The first ($\dot{a}_{\ell,\, \rm GC}$) arises from the movement of the pulsar in the potential of the
cluster: different positions in the cluster will generally 
have a different $a_{\ell,\, \rm GC}$; the movement of the pulsar
from one to the other will therefore cause a variation of this quantity.
The second contribution to $\dot{a}_\ell$ is due to the gravity of nearby stars;
this is more significant for denser clusters.

\cite{fck+03} detected the second spin frequency derivative for
only one pulsar, 47~Tuc~H
($\ddot{f} \, = \, 1.6 \, \pm \, 0.2 \, \times \, 10^{-25}\, \rm Hz \, s^{-2}$).
They then estimated the maximum line-of-sight jerk induced by the motion of the
pulsar in the mean field of the cluster ($\dot{a}_{\ell,\, \rm GC,\, max}$)
and the corresponding $\ddot{f}$ ($\ddot{f}_{\rm max}$)
using a slightly modified version of Equation 4.3 in \cite{phi93}:
\begin{equation}
\frac{\dot{a}_{\ell,\, \rm GC,\, max} (0)}{c}\, = \, -\frac{\ddot{f}_{\rm max}}{f} \, = \, - \frac{4 \pi}{3}G \rho(0)\, \frac{v_{\ell, \rm max}}{c},
\label{eq:dot_a_l}
\end{equation}
where $v_{\ell, \rm max}$ is the maximum velocity of a pulsar relative to
the cluster, in this case assumed to be moving along the line of sight
through the centre of the cluster. If $v_{\ell, \rm max}$ is positive
(i.e., the pulsar is moving away from us), then $\dot{a}_{\ell,\, \rm GC,\, max} (0)$
is negative, and vice-versa.

\cite{fck+03} used $\sigma_0 \, \sim \, 13 \, \rm km \, s^{-1}$
as an estimate of $v_{\ell, \rm max}$. The $\ddot{f}$ of 47 Tuc H is much larger than
the resulting $\ddot{f}_{\rm max}$, from this they concluded
that this system is being perturbed by a nearby stellar companion.

However, that estimate of $\dot{a}_{\ell,\, \rm GC,\, max}$
(and $\ddot{f}_{\rm max}$) is not very precise:
First, because Eq.~\ref{eq:dot_a_l} applies only to the centre of the
cluster; second because,
as we have seen in Section~\ref{sec:proper_motions}, individual pulsars
can have velocities along any direction that are almost twice as large
as $\sigma_0$. Owing to our larger timing baseline $T$,
we are now able to measure $\dot{a}_\ell$
precisely for almost all MSPs in 47~Tuc  (see Tables~\ref{tab:isolated}
to \ref{tab:MSP_BW}); this improvement in measurements of $\dot{a}_\ell$ must be
matched by an improvement in the prediction of
$\dot{a}_{\ell, \, \rm GC,\, max}$.

This prediction is obtained from
the gradient of $a_{\ell,\, \rm GC}$ along the direction of the line of sight $l \equiv \ell r_c$
where it reaches a maximum, at $l = \ell = 0$ (i.e. in the plane $\Pi$, 
defined in section~\ref{sec:pdots})
and then multiplying it by $v_{\ell, \, \rm max}$. Near this plane
$\ell$ is small, so $x = \sqrt{\ell^2 + x_{\perp}^2} \simeq x_{\perp}$
is basically independent of $\ell$. In that case,
the line-of-sight accelerations can be derived trivially from Eq.~\ref{eq:accel2}
multiplied by the projection factor $\ell/x_{\perp}$:
\begin{equation}
\label{eq:accel0}
a_{\ell, \rm GC}(x_{\perp}) = \frac{9 \sigma^2_{\mu, 0}\, d}{\theta_c} \frac{\ell}{x_{\perp}^3}
\left( \frac{x_{\perp}}{\sqrt{1+x_{\perp}^2}}  - \sinh^{-1}x_{\perp} \right),
\end{equation}
Being proportional to $\ell$ in the vicinity of $\Pi$, these line-of-sight accelerations
are zero for any object in $\Pi$, so the only non-zero spatial derivative
of $a_{\ell, \rm GC}$ in that plane is along its normal: the direction of the line of sight
$l$. This derivative is trivial since Eq.~\ref{eq:accel0} is linear in $\ell$.
Using ${\rm d} \ell / {\rm d} l = {\rm d} \ell / (d \, \theta_c) {\rm d} \ell = 1 /(d \, \theta_c)$,
we obtain, for $\ell \, = \, 0$ (not forgetting $v_{\ell, \, \rm max}$):
\begin{equation}
\dot{a}_{\ell,\, \rm GC, \rm max} (x_\perp)
=  \frac{9 \sigma_{\mu, 0}^2}{\theta_c^2} \frac{1}{x_{\perp}^3}
\left( \frac{x_{\perp}}{\sqrt{1 + x^2_{\perp}}} - 
 \sinh^{-1}x_{\perp}
 \right) v_{\ell, \, \rm max}.
\label{eq:a-dot}
\end{equation} 
% Comparing Eq.~\ref{eq:a-dot} and with Eq.~\ref{eq:accel2}, we obtain:
% \begin{equation}
% \dot{a}_{l,\, \rm GC, \rm max} (x_\perp) = 
% \frac{a_{\rm GC}(x_{\perp})}{d \, \theta_{\perp}} v_{l, \, \rm max}.
% \end{equation}
For the line of sight going through the centre ($x_{\perp} = \theta_{\perp} \, = \, 0$), 
Eq.~\ref{eq:a-dot} cannot be evaluated directly, but the limit
of the terms with $x_{\perp}$ is $-1/3$. Thus, in that limit
we recover the result of Eq.~\ref{eq:dot_a_l} (with the central density from
Eq.~\ref{eq:central_density}) for the  most extreme $\dot{a}_{\ell,\, \rm GC, \rm max}$:
\begin{equation}
\dot{a}_{\ell,\, \rm GC, \rm max}(0)\, = \, - \frac{3 \sigma^2_{\mu, 0}}{\theta^2_c} v_{\ell, \, \rm max}.
\label{eq:central}
\end{equation}
Apart from $v_{\ell, \, \rm max}$ these predictions for $\dot{a}_{\ell,\, \rm GC, \rm max}$
depend only on angular measurements.
In our calculations, we used $v_{\ell, \, \rm max} \, = \, v_e$,
the velocity envelope derived in section~\ref{sec:proper_motions}.

The comparison between this prediction and the measured jerks is displayed graphically in
Fig.~\ref{fig:accel-dot}. The red dots and red vertical errorbars
depict the measurement of the jerks and associated uncertainties.
We also plot (in solid black lines) the $\dot{a}_{\ell, \, \rm GC,\, max}$
predicted for each line of sight by our cluster model.
Most pulsars have line-of-sight jerks that are smaller than
our estimate of $\dot{a}_{\ell,\, \rm GC, \, max}$ for their lines of sight;
such jerks can therefore be attributed to the movement of the pulsars
in the mean field of the cluster.

However, a few stand out. For 47~Tuc~H, the observed $\ddot{f}$ is consistent
with that reported in 2003, but 10 times as precise: 
$\ddot{f} \, = \, 1.60 \, \pm \, 0.02 \, \times \, 10^{-25}\, \rm Hz \, s^{-2}$;
the corresponding line-of-sight jerk
($\dot{a}_\ell \, = \,-1.545(22) \, \times \, 10^{-19}\, \rm m \, s^{-3}$)
is much larger (in absolute terms)
than $|\dot{a}_{\ell, \, \rm GC,\, \rm max}|$ for that pulsar's line of sight
(or any in the cluster), so we come to the
same conclusion as \cite{fck+03}: this system must have a nearby companion.
We can now see that this is also true for 47~Tuc~U and J.
For 47~Tuc~M and aa the observed jerks are only $\sim$1 $\sigma$ away from
the $\dot{a}_{\ell, \, \rm GC, \rm max}$ for their lines of sight.

The systems with line-of-sight jerks larger than
the maximum cluster mean-field expectation lie at distances from the core
of about 1\arcmin, not near the centre. Given the larger density
of stars near the centre one might expect that larger jerks would occur there,
however the predicted mean-field jerks are also larger near the centre.
The numerical simulations  presented in \cite{prf+17} suggest
that the probability for a jerk from a nearby companion to be significantly
larger than the cluster mean-field contribution
is relatively flat with distance from the centre of the cluster.
This means we are likely to find systems like 47 Tuc H at any distance from
the centre.

\subsection{Third spin frequency derivative}

Unlike for the lower spin frequency derivatives, the third and higher
spin frequency derivatives, $f^{(n)}$, can only be caused by the gravitational field of
nearby objects.

The idea that 47~Tuc~H is being influenced by a nearby stellar companion
is supported by the fact that it is the pulsar in the cluster for which the
measurement of $f^{(3)}$ is most significant,
$f^{(3)}\, = \, (3.8\, \pm \, 1.7)\, \times \, 10^{-35}\, \rm Hz\, s^{-3}$,
a 2.3-$\sigma$ ``detection''.
Most of the black widow systems also appear to have a non-zero
$f^{(3)}$, but in no case is the measurement more significant than
2-$\sigma$. For the other main candidate for a stellar
companion, 47~Tuc~U, we measure
$f^{(3)}\, = \, (-1.0\, \pm \, 1.3)\, \times \, 10^{-35}\, \rm Hz\, s^{-3}$;
we therefore did not fit for this parameter in the derivation of its
timing solution (Table~\ref{tab:MSP_WD}).

Since the uncertainty in the measurement of $f^{(3)}$
scales with $T^{-9/2}$, continued timing will improve these measurements very quickly.
The rate of improvement will be even faster for higher frequency
derivatives. The measurement of five such derivatives  allows a unique determination of the five
Keplerian orbital parameters \citep{jr97}; we would then know
whether the 47~Tuc~H system and the nearby star are bound or not.

\section{Orbital period derivatives}
\label{sec:accelerations}

\begin{figure*}
	\includegraphics[width=0.7\textwidth]{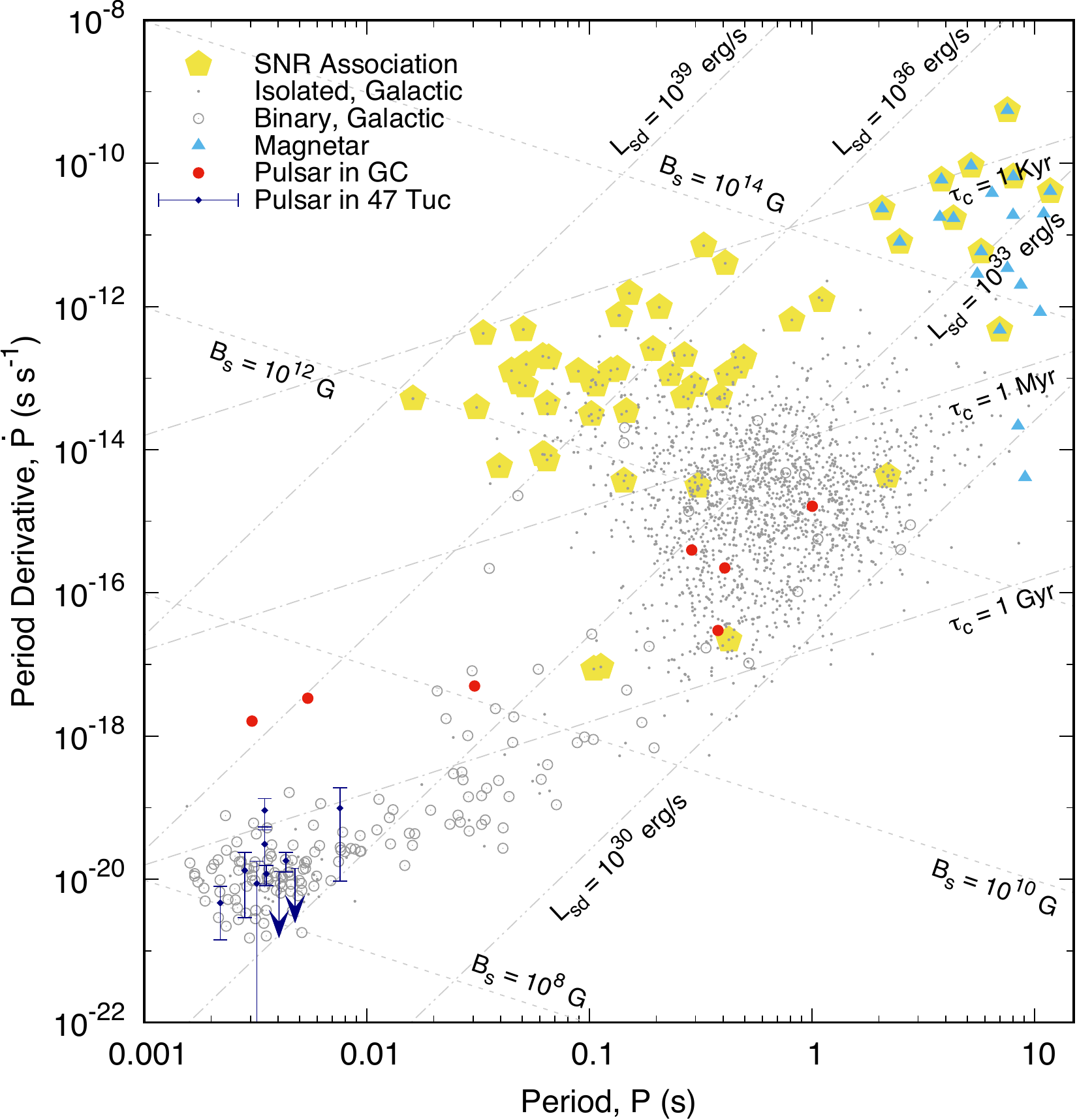}
    \caption{Period - period derivative plot for the pulsars in the ATNF Pulsar Catalogue
    \citep{mhth05}.
    The newly derived period derivatives for the MSPs in 47~Tuc (in dark blue) place them in
    the same region of the diagram where the majority of Galactic MSPs occur,
    i.e., they appear to be normal millisecond pulsars. Based on this sample, we conclude
    that 47~Tuc does not appear to have young pulsars like some seen in some other
    globular clusters (red dots).}
\label{fig:P-Pdot}
\end{figure*}

One of the quantities that benefits most from prolonged timing is the measurement of the
variation of the orbital period, $\dot{P}_b$. For most of the eclipsing binaries like
47~Tuc~V and W (see Paper I) and 47~Tuc~J and O (see Section~\ref{sec:spiders}),
there are unpredictable variations in the orbital period with time, similar to those
observed for other eclipsing binaries in the Galaxy (e.g., \citealt{svf+16});
in these cases we need many
orbital frequency derivatives to describe the evolution of orbital phase with time.
For the remaining binary pulsars -
the MSP-WD systems (47~Tuc~E, H, Q, S, T, U, X and Y) and two of the black widow systems
(47~Tuc~I and R) the phase evolution of the orbit can be described with a
period and a period derivative only. 

\subsection{Measurements of accelerations}

If the orbital period, $P_{\rm b}$, in the reference frame of the binary is stable,
then we will not be able to measure orbital frequency derivatives
higher than $1^{\rm st}$ order (unless the system is in a triple - in which case the
effects will be much more obvious in the spin frequency derivatives).
At the Earth, the observed orbital period
derivative will then be given by \citep{dt91}:
\begin{equation}
\frac{\dot{P}_{\rm b, \rm obs}}{P_{\rm b}} \, = \, \frac{\dot{P}_{\rm b, \rm int}}{P_{\rm b}} \, + \, \frac{\mu^2 d}{c} + \frac{a_{\ell, \, \rm GC}}{c} + \frac{a}{c},
\label{eq:pb_dot}
\end{equation}
where all parameters are as in equation~(\ref{eq:p_dot}), except that
$\dot{P}_{\rm b, \rm obs}$ is the observed orbital period derivative and
$\dot{P}_{\rm b\, \rm int}$ is the intrinsic orbital period derivative.
For the MSP-WD systems, the intrinsic variation of the orbital period,
$\dot{P}_{\rm b, \rm int}$, should be dominated by energy loss
due to the emission of gravitational waves. This is expected to be a very small
quantity: for the MSP-WD system with the shortest orbital period,
47~Tuc~U ($P_{\rm b} \, = \, 0.42911 \rm \,d$), the  orbital decay expected is
$- 1.36 \, \times \, 10^{-14} \rm s\, s^{-1}$ (this
assuming that the MSP has a mass of $1.4\, M_{\odot}$ and an orbital
inclination $i = 90^\circ$), which is a factor of 2 smaller than the
current measurement uncertainty for the $\dot{P}_{\rm b, \rm obs}$ for
that pulsar. The cases of 47~Tuc~I and R are discussed in detail in
Section~\ref{sec:spiders}.

Re-writing equation~(\ref{eq:pb_dot}), and ignoring the intrinsic term,
we can, for each pulsar, calculate the cluster
acceleration, since the remaining terms are also known, in particular
the proper motion (see Section~\ref{sec:proper_motions}):
\begin{equation}
a_{\ell,\, \rm GC}\, = \, \frac{\dot{P}_{\rm b,\, \rm obs}}{P_{\rm b}}c - \mu^2 d - a.
\label{eq:a_l}
\end{equation}
These accelerations are presented in
Tables~\ref{tab:MSP_WD} and \ref{tab:MSP_BW}, and depicted graphically 
as the vertical red error bars in Fig.~\ref{fig:accel}. Like
the values of $\dot{P}_{\rm obs}/P$, they
represent important constraints on the dynamics of the cluster.
As we can see in Fig.~\ref{fig:accel}, the
accelerations for 47~Tuc~S, E and U can (just about) 
be accounted for by the mass model for the cluster described in
Section~\ref{sec:jerks} with a distance of 4.69 kpc.
With the kinematic distance (4.15 kpc,
represented by the dotted line in Fig.~\ref{fig:accel}),
this model cannot account for these accelerations.

We conclude therefore that our acceleration measurements are
not compatible with a $d$ significantly smaller than
4.69 kpc, in agreement with most published distance estimates;
they appear to be incompatible with the kinematic distance
of 4.15 kpc. A more robust probabilistic estimate
of the cluster distance most favoured by our measurements
will be presented elsewhere.

\subsection{Intrinsic spin period derivatives}

As we can see from Fig.~\ref{fig:accel}, the measured
values of $a_{\ell, \, \rm GC}$ tend to be similar,
but slightly smaller than $a_{\ell, \rm max}$.
The difference, as can be seen in equation~(\ref{eq:almax}),
is due to the contribution from $\dot{P}_{\rm int}$.
The values of $\dot{P}_{\rm int}$ can be obtained directly
from the observables  by subtracting equation~(\ref{eq:pb_dot})
from equation~(\ref{eq:p_dot}) and re-arranging the terms
(and taking into account the fact that the $\dot{P}_{\rm b, \rm int}$
are small):
\begin{equation}
\dot{P}_{\rm int} = \dot{P}_{\rm obs} - \frac{\dot{P}_{\rm b, \rm obs}}{P_{\rm b}} P.
\label{eq:pdot_int}
\end{equation}
The intrinsic values of $\dot{P}$ are
 presented in Tables~\ref{tab:MSP_WD} and \ref{tab:MSP_BW}.
For most of the other pulsars, less constraining upper limits for
$\dot{P}_{\rm int}$ were 
derived assuming the largest possible negative value of $a_{\rm GC}$
for the line of sight of the pulsar (see \citealt{fcl+01} and \citealt{fck+03}).

Although the estimates of $\dot{P}_{\rm int}$ are not measured with
high significance (only a couple of cases, 47~Tuc~E and U, are measured
with 3-$\sigma$ significance), they already allow a comparison
with the MSPs in the Galaxy. Putting the limits in a $P$ - $\dot{P}$ diagram
(Figure~\ref{fig:P-Pdot}), we see that these pulsars have characteristics
(spin-down energy, magnetic fields at the poles, characteristic ages) 
similar to the majority of MSPs in the disk of the Galaxy. They are very different from
some of the ``young'' globular cluster pulsars (depicted in red),
for which the $\dot{P}_{\rm obs}$ are too large to be explained by
cluster accelerations (for a discussion, see e.g., \citealt{faa+11},
\citealt{jgk+13}, \citealt{vf14} and references therein).

The relatively large relative uncertainties
of $\dot{P}_{\rm int}$ imply that the derived magnetic fields, spin-down
powers and characteristic ages of these pulsars still have large uncertainties.
In what follows we calculate explicitly the characteristic
ages, $\tau_c$, or lower limits on them, these are also presented in
Tables~\ref{tab:MSP_WD} and \ref{tab:MSP_BW} and were calculated using
$\tau_c = P / (2 \dot{P}_{\rm int})$. The idea is to compare them with
the total ages ($\tau_o$) estimated for the
WD companions that have been detected by the \emph{HST}
\citep{egh+01,egc+02,rbh+15,cpf+15}.
These estimates agree, i.e., we find
no case where $\tau_c \, << \, \tau_o$.
A similar comparison was done in \cite{rbh+15} and \cite{cpf+15}
using preliminary numbers from our timing program.
It is interesting to note that the two apparently oldest WD companions,
47~Tuc~Q and Y, are those that have the largest lower limits for $\tau_c$.

\subsubsection{47 Tuc Q}

For 47~Tuc~Q, $\dot{P}_{\rm int} \, = \, (-5.8 \, \pm \, 9.3) \, \times 10^{-21}\, \rm s \, s^{-1}$.
This means that we cannot specify an upper limit for 
$\tau_c$, since
$\dot{P}_{\rm int}$ could be very small. Its 2-$\sigma$
upper limit, $1.28 \, \times 10^{-20}\, \rm s \, s^{-1}$, implies
a minimum $\tau_c$ of 5.0 Gyr.

For a variety of reasons, the $\tau_o$ for the WD companion
of this pulsar is highly uncertain: the cooling age ranges from 0.3 to 5 Gyr
(this value depends very sensitively on the mass of the WD), plus
$\sim\, 1\, \rm Gyr$ for the proto-WD phase \citep{rbh+15}. We thus find that
an age close to 6 Gyr is preferred for this system.

\subsubsection{47 Tuc S}

For 47~Tuc~S, $\dot{P}_{\rm int} \, = \, (1.3 \, \pm \, 1.0) \, \times 10^{-20}\, \rm s \, s^{-1}$.
Again, no reliable upper age can be derived,
but a lower limit for $\tau_c$ of 
1.3 Gyr can be derived from the 2-$\sigma$ upper limit of
$\dot{P}_{\rm int}$. The cooling age ranges from 0.1 to 0.4 Gyr, to this we should add
up to 0.4 Gyr for the time the companion spent as a
proto-WD \citep{rbh+15}. This suggests that $\tau_o$
is not larger than 0.8 Gyr.
This is fine since $\tau_c$ assumes a starting spin period
that is much shorter than the present spin period. This is clearly
not the case for most MSPs, particularly those with shorter
spin periods, meaning that the real age will
generally be smaller than $\tau_c$.

\subsubsection{47 Tuc T}

For 47~Tuc~T, the timing constraints are not so precise and
we get $\dot{P}_{\rm int} \, = \, (9.9 \, \pm \, 8.9) \, \times 10^{-20}\, \rm s \, s^{-1}$.
This implies a 2-$\sigma$ lower limit $\tau_c \, > \, 0.43\,$Gyr.
The estimated $\tau_o$ is from 0.1 to 0.8 Gyr \citep{rbh+15}, consistent
with $\tau_c$.

\subsubsection{47 Tuc U}

For 47~Tuc~U,
$\dot{P}_{\rm int} \, = \, (1.82 \, \pm \, 0.55) \, \times 10^{-20}\, \rm s \, s^{-1}$,
this means we can establish solid lower {\em and} upper limits for $\tau_c$ from the
2-$\sigma$ upper and lower limits of $\dot{P}_{\rm int}$:
$2.4 \, < \, \tau_c \, < \,9.4 \, $Gyr. For the WD companion \cite{rbh+15}
derive a $\tau_o$ between 1.6 - 2.1 Gyr, slightly lower than $\tau_c$.
As for 47~Tuc~S, this is fine since $\tau_c$ represents an
upper limit on the age that assumes a very small initial spin period.

\subsubsection{47 Tuc Y}

For this pulsar 
$\dot{P}_{\rm int} \, = \, (4.7 \, \pm \, 3.3) \, \times 10^{-21}\, \rm s \, s^{-1}$,
from the 2-$\sigma$ upper limit of $\dot{P}_{\rm int}$ we derive a lower
limit for $\tau_c$ of 3.1 Gyr. \cite{rbh+15} derive
a $\tau_o$ between 3.1 to 3.9 Gyr, in agreement with $\tau_c$.

\begin{figure*}
  \includegraphics[width=0.95\columnwidth]{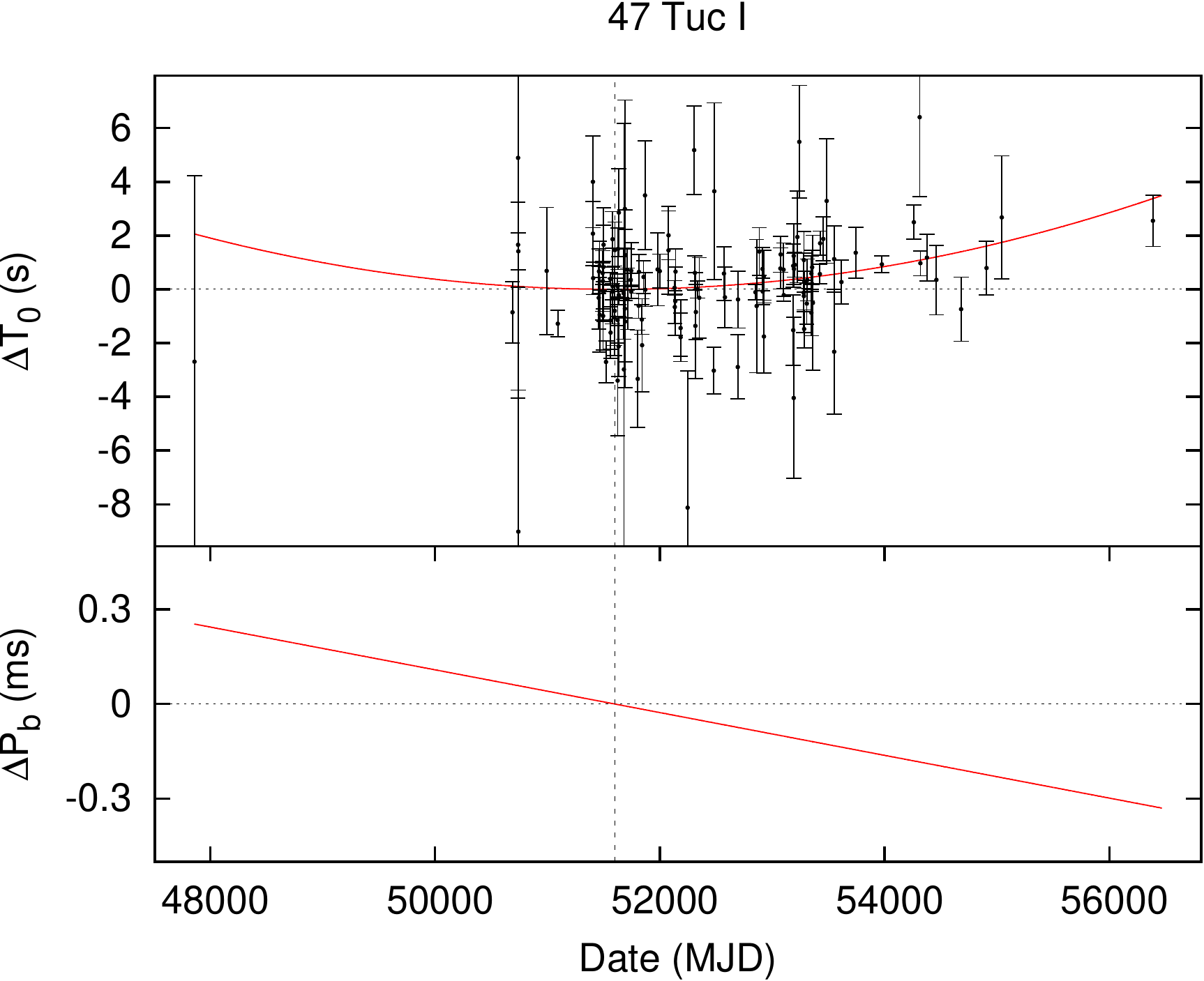}
  \qquad 
    \includegraphics[width=0.95\columnwidth]{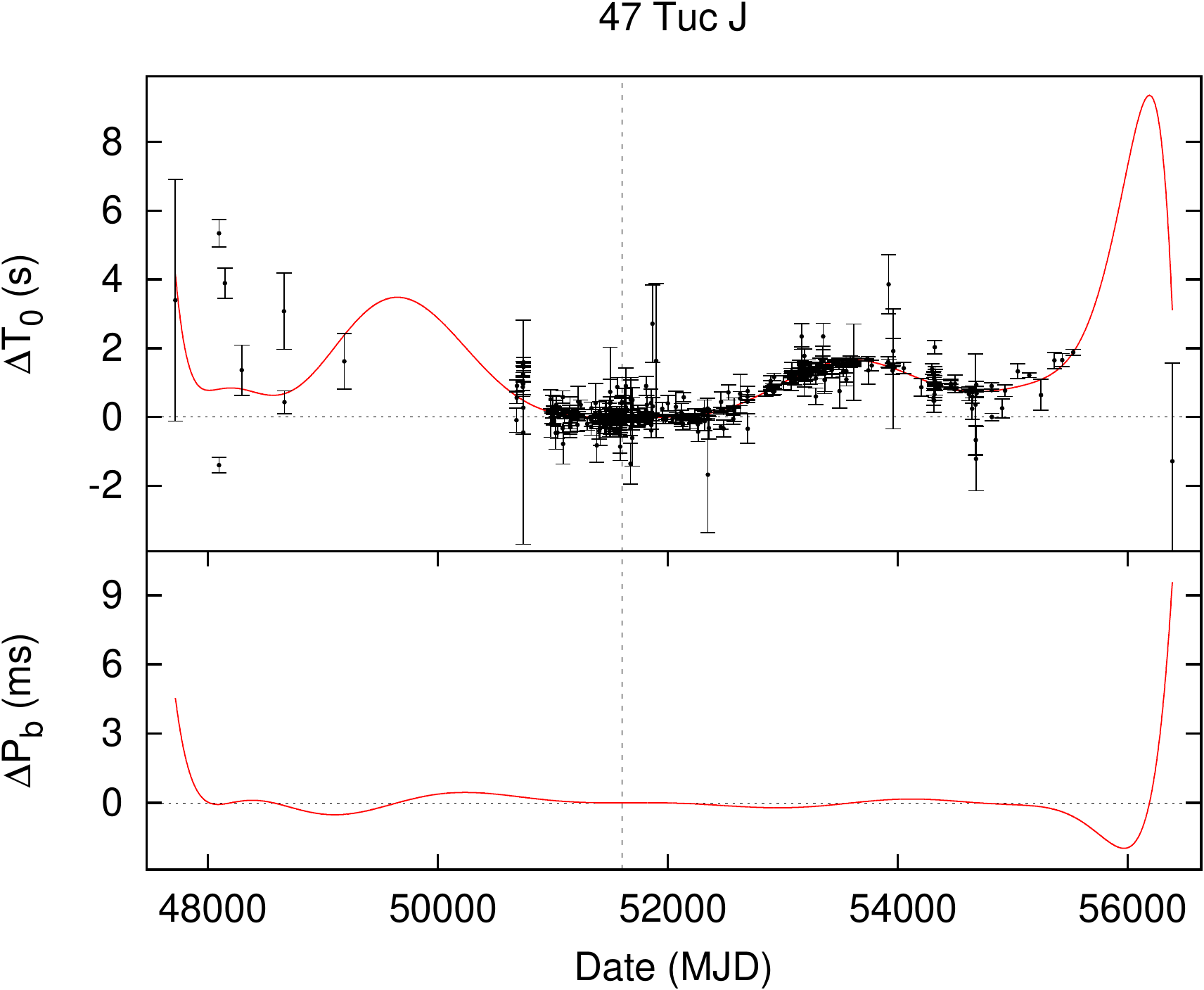}
    \\
   \vskip 0.7cm
  \includegraphics[width=0.95\columnwidth]{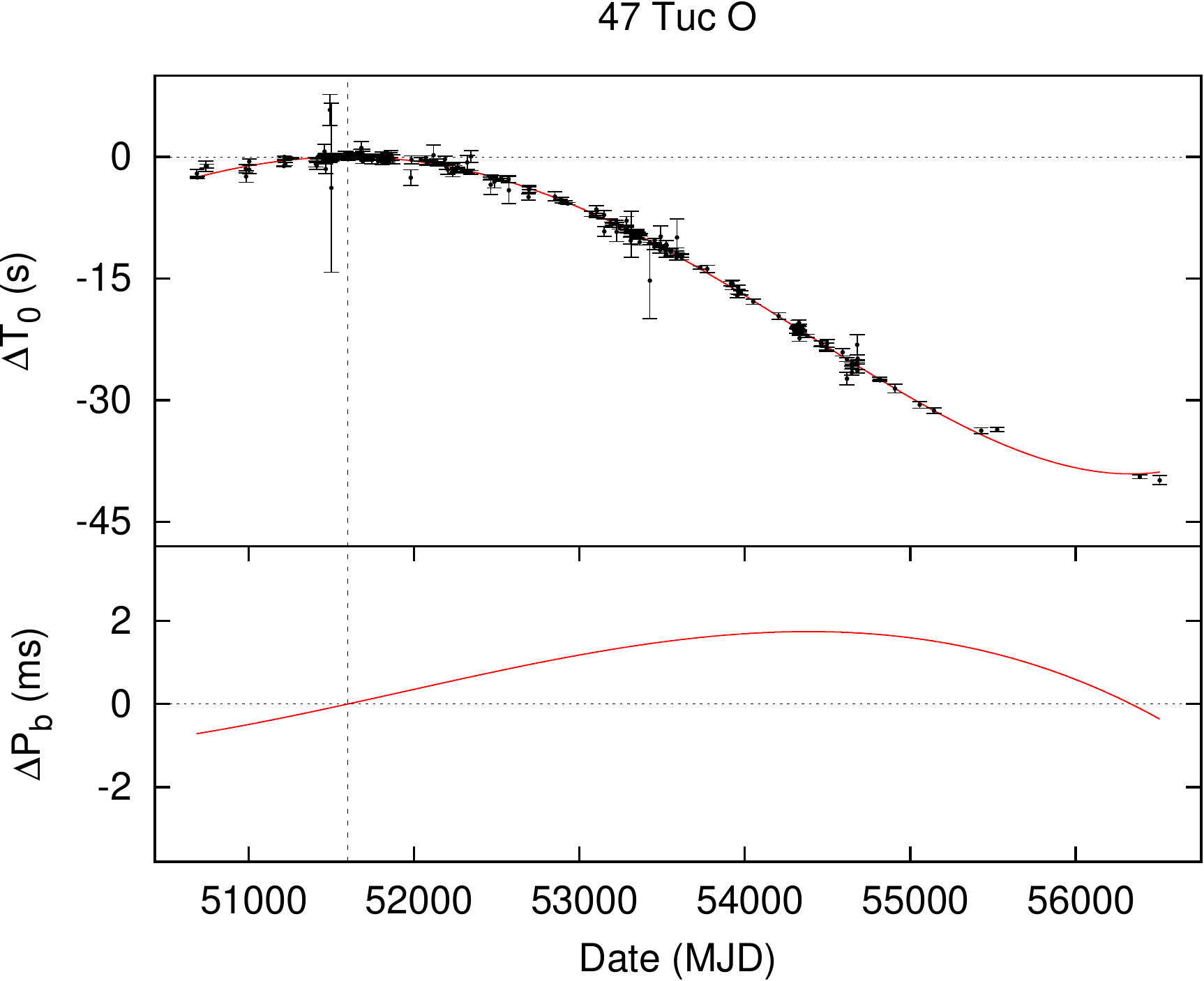}
  \qquad
  \includegraphics[width=0.95 \columnwidth]{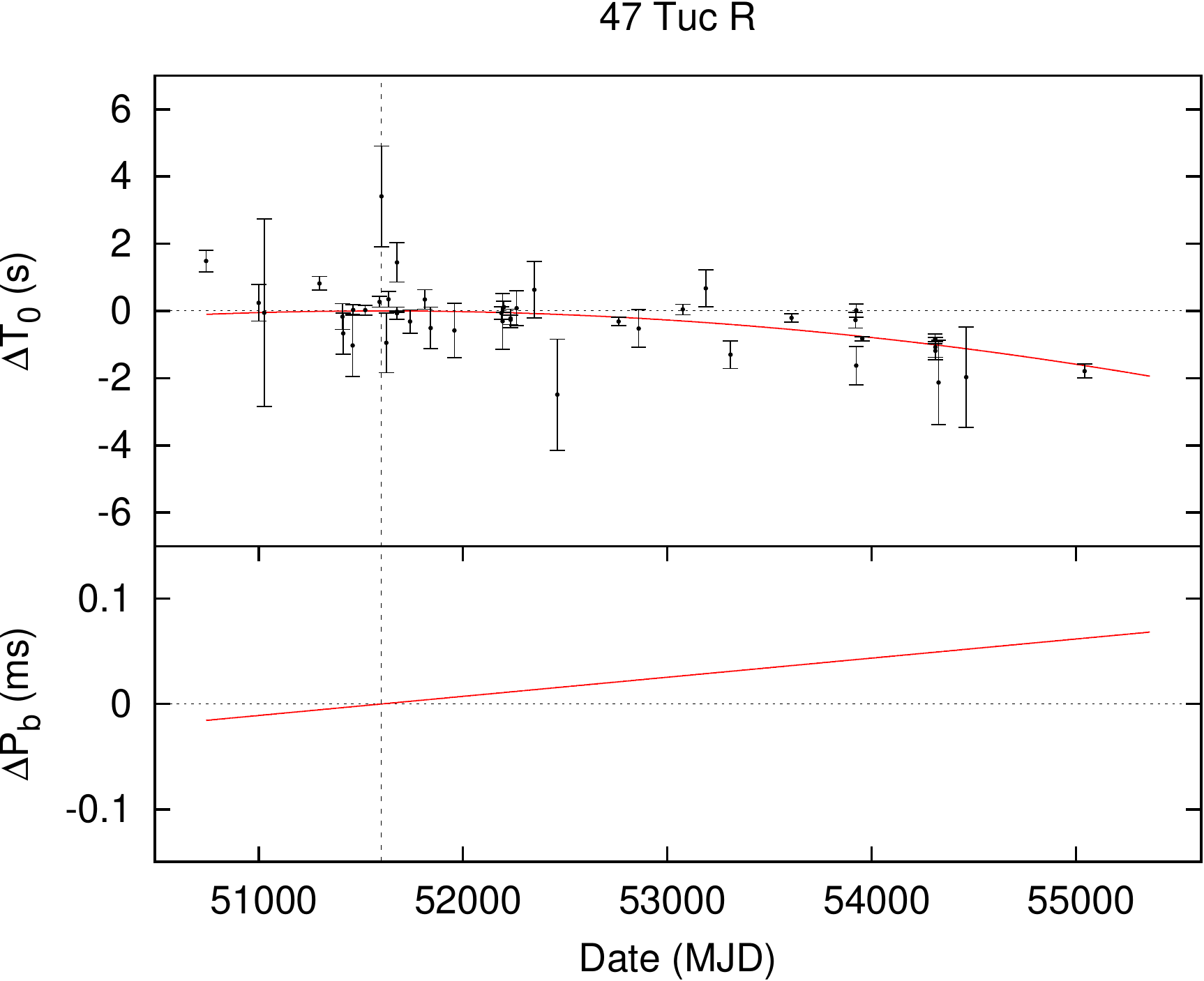}
  \caption{ Orbital variability of the four black widows 47 Tuc I, J, O and R. In all the plots,
    the vertical dotted line indicates the passage through ascending node
    closest to the reference epoch, MJD = 51600 (see Table~\ref{tab:MSP_BW}).
    In the upper panels of each plot the vertical axis
    represents $\Delta\, T_0$ in the BTX model and $\Delta\, T_{\rm asc}$
    in the ELL1 model. For the lower panels of each plot the vertical axis
    represents the corresponding change in $P_b$. For the orbit closest to
    the reference epoch $\Delta\, T_0 \, \equiv\, \Delta\, T_{\rm asc} \, = \,0$
    and $\Delta P_{\rm b}\, = \, 0$. In the case of 47 Tuc J, given the large number of
    orbital frequency derivatives, the model predicts large orbital phase swings
    outside the timing baseline; this is not an accurate prediction
    of the system's orbital phase evolution.}
  \label{fig:orb_var}
\end{figure*}

\section{Black widows}
\label{sec:spiders}

There are five black widow pulsars known in 47~Tuc, namely 47~Tuc~I, J, O, and R
(discussed below) and 47~Tuc~P, studied in detail in Paper I.
Black widow binary systems are mostly defined by their short
orbital periods, small ($< \, 0.05\, M_{\odot}$) companion masses,
and the detectability of radio eclipses, although not for every system, 
while on the other hand redback pulsars have more massive  %removed "(RB)", since it wasn't used again; C.Heinke
companions ($> \, 0.1\, M_{\odot}$) and always display eclipses (see e.g. 
\citealt{Freire2005,Roberts13} for reviews). The two redbacks in 47 Tuc,
i.e. pulsars V and W, were studied in detail in Paper I. 

Both types of systems are known for their
orbital variability: the orbital period (and sometimes the projected
semi-major axis) change unpredictably with time, as seen 
in long-term timing of some black widow systems (see, e.g., \citealt{svf+16}).
Such variability requires the use of the BTX
orbital model (D. Nice, unpublished; \url{http://tempo.sourceforge.net/}),
which allows a
description of the orbital behaviour using multiple orbital frequency derivatives.
In order to characterise the orbital variability we also use a method described
in Paper I (Section 5.2.1) and in \cite{svf+16}, where we make multiple measurements
of the time of ascending node, $T_{\rm asc}$ as a function of time.

The results can be seen in Fig. \ref{fig:orb_var},
where we depict the orbital phase (and orbital period) evolution with time, and compare
the observed $T_{\rm asc}$ with the expectation based on the timing models in
Table~\ref{tab:MSP_BW}. The observed orbital phase variations are relatively smooth
and are well described by the global models. 

\subsection{Black widows with large orbital variability}

As mentioned before, two of the black widow systems, 47~Tuc~J and O, display
this characteristic variability. For 47 Tuc J, the orbital period appears to
vary by a fraction of 1 ms in a quasi-sinusoidal fashion.
Twelve orbital frequency derivatives were thus necessary to correctly
model this behaviour within the timing baseline. The orbital period of 47 Tuc O, instead,
shows a constant increase until MJD $\sim 54300$, then a decrease later. In this case, only 
three orbital frequency derivatives were necessary to model the variability. This is
in part due to the fact that the timing baseline for this pulsar is  shorter
than for 47~Tuc~J. These variations are not caused by any nearby objects, as the motion
of the system would be obvious in variations of the observed pulse period.

It is important to note that, for these pulsars, the BTX models are only valid in the
time span covered by the available data, i.e. they do not have predictive power and cannot accurately
describe the orbital phase evolution outside the timing baseline.

\subsection{Black widows with small orbital variability}

Even with the long timing baseline being considered in this paper,
the orbits of 47 Tuc I and R
can be described without the need of introducing any orbital frequency derivatives
higher than the first (which in these cases we report as
$\dot{P}_{\rm  b, obs}$). This could be due, to some extent, to lack of timing precision.
Looking at Fig. \ref{fig:orb_var}, we can see that the oscillations in $\Delta T_{\rm asc}$
for 47~Tuc~J are quite small compared to the dispersion of the data points
observed in 47~Tuc~I and R.
If the latter could be timed with the same precision as 47~Tuc~J, it is possible that
subtle oscillations in $T_{\rm asc}$ (such as those observed for 47~Tuc~J) would become
detectable.

However, the values of $\dot{P}_{\rm b, \rm obs}/P_{\rm b}$ for these two systems are
remarkably similar to their $\dot{P}_{\rm obs}/P$
(see Fig.~\ref{fig:accel}); they are even slightly smaller as one would expect from
a positive intrinsic spin period derivative $\dot{P}_{\rm int}$:
for 47~Tuc~I, we obtain  $\dot{P}_{\rm int} \, = \, (9.2 \, \pm \, 4.3) \, \times 10^{-20}\, \rm s \, s^{-1}$,
and for 47~Tuc~R $\dot{P}_{\rm int} \, = \, (3.1 \, \pm \, 2.2) \, \times 10^{-20}\, \rm s \, s^{-1}$,
values that are similar to those of the remaining pulsars.
This makes it likely that, as in the case of
$\dot{P}_{\rm obs}/P$, the $\dot{P}_{\rm b, \rm obs}/P_{\rm b}$ observed in these two systems is
mostly caused by the acceleration of the pulsars in the field of the globular cluster, 
$a_{\ell,\, \rm GC}$.

% \begin{figure}
%   \includegraphics[width=\columnwidth]{47TucI_variability.pdf}
%     \caption{Orbital Variability of 47 Tuc I. In this and the following plots,
%     the vertical dotted line indicates the passage through ascending node
%     closest to the reference epoch, MJD = 51600 (see Table~\ref{tab:MSP_BW}).
%     The vertical axis
%     represents $\Delta\, T_0$ in the BTX model and $\Delta\, T_{\rm asc}$
%     in the ELL1 model. For the orbit closest to the reference epoch
%     $\Delta\, T_0 \, \equiv\, \Delta\, T_{\rm asc} \, = \,0$ and $\Delta P_{\rm b}\, = \, 0$.
%     }
%     \label{fig:orb_var_47TucI}

%     \includegraphics[width=\columnwidth]{47TucJ_variability.pdf}
%     \caption{Orbital Variability of 47 Tuc J. Given the large number of orbital frequency
%     derivatives, the model predicts large orbital phase swings
%     outside the timing baseline; this is not an accurate prediction
%     of the system's orbital phase evolution.}
%     \label{fig:orb_var_47TucJ}
% \end{figure}

These systems have such short orbital periods that, despite
the small companion masses, we must take into account an intrinsic
variation of the orbital period caused by the emission of gravitational waves.
In this case, equation~(\ref{eq:pdot_int}) becomes:
\begin{equation}
\dot{P}_{\rm int} = \dot{P}_{\rm obs} - \frac{\dot{P}_{\rm b, \rm obs} - \dot{P}_{\rm b, \rm int}}{P_{\rm b}} P.
\label{eq:pdot_int_full}
\end{equation}
Assuming a pulsar mass of $1.4\, M_{\odot}$ and an orbital inclination
of $60^\circ$ for both pulsars and using the equations in \cite{dt91},
we obtain for the orbital decay 
$\dot{P}_{\rm b, \rm int}\, = \, -4.8 \, \times\, 10^{-15}\, \rm s \, s^{-1}$
for 47~Tuc~I
and $\dot{P}_{\rm b, \rm int}\, = \, -7.6 \, \times\, 10^{-14}\, \rm s \, s^{-1}$
for 47~Tuc~R. Inserting these terms in equation~\ref{eq:pdot_int_full}, we
obtain even smaller intrinsic spin period derivatives:
$\dot{P}_{\rm int} \, = \, (7.8 \, \pm \, 4.3) \, \times 10^{-20}\, \rm s \, s^{-1}$
for 47~Tuc~I and 
$\dot{P}_{\rm int} \, = \, (-1.6 \, \pm \, 2.2) \, \times 10^{-20}\, \rm s \, s^{-1}$
for 47~Tuc~R, implying lower limits on the characteristic ages of both systems
of 0.33 and 2.0 Gyr respectively.
This means that the
agreement between the cluster acceleration and the observed 
orbital period derivative is even more precise when we take  the gravitational
wave emission into account.
We conclude, provisionally, that the black widow systems
come in two flavours, with and without random orbital variability.

\section{New detections of the rate of advance of periastron}
\label{sec:binary_masses}

Another measurement that benefits greatly from a much extended timing
baseline is the rate of advance of periastron, $\dot{\omega}$.
For a binary system consisting of two point masses (a reasonable
approximation for the MSP-WD binaries in 47~Tuc), this 
is solely an effect of relativistic gravity. In general relativity, and to leading
post-Newtonian order, it depends only on the Keplerian
parameters and the total mass of the system $M$, in solar masses
(\citealt{rob38,tw82}):
\begin{equation} \label{eq:omdot}
  \dot{\omega}_{\rm GR} = 3 \frac{({\rm T_{\odot}} M)^{2/3}}{1- e^2} 
  \left( \frac{P_{\rm b}}{2\pi } \right)^{-5/3},
\end{equation}
where  ${\rm T}_{\odot}\, \equiv \, G M_{\odot} c^{-3} \, = \, 4.9254909476412675\, \mu
$s is a solar mass ($M_{\odot}$) in time units, $c$ is the speed of light and $G$ is
Newton's gravitational constant.

To measure this effect, we need a system with a 
significant orbital eccentricity ($e$), otherwise it is impossible to measure
the longitude of periastron ($\omega$) with sufficient precision to detect
its variation with time. For most MSP-WD systems in the Galaxy,
$e$ is too small for such a measurement to be feasible. In globular clusters,
on the other hand, the stellar density is so high that
binary pulsars are perturbed by close encounters
with other members of the cluster; this will generally increase their orbital
eccentricity \citep{phi92,phi93}. The large eccentricities
of many binaries in GCs has allowed the measurement of 
their $\dot{\omega}$ - and consequently, of the binary masses
(see e.g., \citealt{of16} and references therein). However, the
perturbations (and corresponding increases in $e$) are larger 
for the wider binaries; this implies that, generally, when we
are able to measure $\dot{\omega}$ well, then the wide
orbit makes it hard to measure other relativistic parameters 
(these would be useful for determining the individual
masses of the components of the binary). There are only
two exceptions to date, both products of exchange
interactions located in core-collapsed
clusters (PSR~J1807$-$2500B in NGC 6544, \citealt{lfrj12}, and PSR~B2127+11C
in M15, \citealt{jcj+06}).

Among the known binary pulsars in 47~Tuc, the most eccentric by far
is 47~Tuc~H ($e\, = \, 0.0705585 \pm 0.0000007$), which is
also the second widest known in the cluster ($P_{\rm b} \, = \, 2.3577 \, \rm d$).
This orbital eccentricity is $4-5$ orders of magnitude larger than
observed in MSP-WD systems of similar $P_{\rm b}$ in the Galactic disk.
For this system, \cite{fck+03} measured
$\dot{\omega}\, = \, 0.0658 \, \pm \, 0.0009 ^\circ\, \rm yr^{-1}$
(where the uncertainty is the 1-$\sigma$ error returned by {\sc TEMPO}).
This allowed an estimate of the total mass of the system of
$M \, = \, 1.61 \, \pm \, 0.03 \, M_{\odot}$ (1-$\sigma$).
Our current value is  a factor of five better:
$\dot{\omega}\, = \, 0.06725 \, \pm \, 0.00019 ^\circ \, \rm yr^{-1}$; this implies
$M \, = \, 1.665 \, \pm \, 0.007 \, M_{\odot}$ (1-$\sigma$).
No other relativistic orbital effects are detectable, so it is not possible to
determine the individual masses in this binary. However,
combining this constraint with the  constraint $\sin i \leq 1$,
we obtain $M_{\rm p} \, <\, 1.49 \, M_{\odot}$ and $M_{\rm c} \, >\, 0.175 \, M_{\odot}$, for the mass of the pulsar and of the companion, respectively.

Although much lower, the eccentricities of most of the MSP-WD systems in 47~Tuc
are also orders of magnitude larger than observed among MSP-WD systems
with similar orbital periods in
the Galactic disk. Because of this, we have made significant ($> \, 3 \, \sigma$)
detections of $\dot{\omega}$ in 3 other systems: 47~Tuc~E
($\dot{\omega}\, = \, 0.090 \, \pm \, 0.016 ^\circ \rm yr^{-1}$,
$M \, = \, 2.3 \, \pm \, 0.7 \, M_{\odot}$),
47~Tuc~S ($\dot{\omega}\, = \, 0.311 \, \pm \, 0.075 ^\circ \rm yr^{-1}$,
$M \, = \, 3.1 \, \pm \, 1.1 \, M_{\odot}$) and 
47~Tuc~U ($\dot{\omega}\, = \, 1.17 \, \pm \, 0.32 ^\circ \rm yr^{-1}$,
$M \, = \, 1.7 \, \pm \, 0.7 \, M_{\odot}$). These measurements
are, however, not yet precise enough to derive any astrophysically
interesting values of the total masses for these systems.
Improving them is important, because if we can determine precise
total masses for these systems, then we will have good estimates
for the masses of these pulsars since
their WD companion masses are relatively well known from
optical photometry \citep{egc+02,rbh+15,cpf+15}.

Another two systems where $\dot{\omega}$ might be detectable in the future are
47~Tuc~Q ($0.46 \, \pm \, 0.22 ^\circ \rm yr^{-1}$) and
47~Tuc~T ($0.30 \, \pm \, 0.28 ^\circ \rm yr^{-1}$), again two
systems for which we have good optical detections of the WD companions.
For the remaining MSP-WD systems (47~Tuc~X and Y), the orbital eccentricities are
too low for a measurement in the foreseeable future.

\section{Discussion}
\label{sec:discussion}

\subsection{What do the pulsars tell us about the cluster?}

The globular cluster 47~Tuc has one of the largest
total stellar interaction rates ($\Gamma$) among clusters in the Milky Way system
\citep{vh87,bhsg13}.
A consequence of this is that, following the many
exchange encounters, many old, ``dead'' neutron stars find
themselves in binaries with main sequence (MS) companions.
Subsequent evolution of these companions causes transfer of
gas to the NSs, i.e., the system becomes a low-mass X-ray
binary (LMXB). After this, the companion typically
becomes a low-mass WD, and the NS becomes a radio MSP.
The large number of MSPs in 47~Tuc can therefore
be understood primarily as a consequence of the large $\Gamma$.

The characteristic ages of the MSPs in 47~Tuc and the optical
ages of their WD companions suggest that these systems have
been forming at a near-constant rate throughout the age of the cluster, i.e., there is
no indication of an early burst of MSP formation (which would
make all pulsars look very old). There are also no signs
of an ongoing burst of MSP formation either - none of the
pulsars in the cluster has a large $\dot{P}_{\rm int}$
that cannot be accounted for by the cluster acceleration model,
none have characteristic ages smaller than about 0.33 Gyr
(the lower limit for the age of 47~Tuc~I). 
In this respect, the situation in 47 Tuc offers a stark contrast to
that observed in some of the core-collapsed clusters,
in particular NGC 6624, where at least three
pulsars (out of the six known in that cluster)
have characteristic ages smaller than 0.2 Gyr \citep{lfrj12}, and in a
particular case (PSR B1820$-$30A) as small
as 25 Myr \citep{faa+11}.

The difference between the populations of these clusters
reflects fundamentally different dynamics. Although
both types of clusters have a similar $\Gamma$,
the interaction rate {\em per binary}, $\gamma$ \citep{vf14} is much
higher in NGC 6624 than in 47~Tuc. The fundamental reason for this
is the fact that NGC~6624 has a collapsed core.

As already discussed in Paper I, the pulsar population in 47 Tuc
has the characteristics one would expect for a
GC with a low $\gamma$: any newly formed LMXBs evolve undisturbed to
their normal outcomes (MSP-WD binaries, black widows and isolated
MSPs, as observed in the Galactic disk). All systems have
large $\tau_c$ the moment they form.  There are no mildly recycled
pulsars - there are currently no companion stars
in GCs massive enough (and evolving fast enough) to
result in mild recycling, as seen for instance in double neutron star
systems and pulsars with massive WD companions in the Galactic disk.
This is the likely reason for the remarkably small range of
spin periods ($1.8 < P < 7.6$ ms) for the pulsars in 47~Tucanae.
Furthermore, the binary systems in 47~Tuc have relatively
small orbital eccentricities compared to what we see in
denser clusters, like Terzan 5 and M28 (for even denser
clusters, binary destruction sets in, but we do see a few
very eccentric survivors). The only "eccentric" binary in
47~Tuc, 47~Tuc~H, might have gained its eccentricity from
an object orbiting it, not from interactions with other stars.

%Also, any LMXBs with orbital periods large enough to only have accretion
%during the giant phase of the companion (case C Roche overflow), which
%could also evolve into mildly recycled binary pulsars,
%are not likely to be stable in these GCs - although possibly stable
%in GCs with much lower stellar densities, like M53.

In clusters with higher $\gamma$, we can find
a higher percentage of isolated pulsars (from the disruption
of MSP-WD systems), mildly recycled - and apparently young - pulsars
(from the disruption of X-ray binaries, which leaves the recycling process
incomplete) and products of {\em secondary} exchange interactions, i.e.,
exchange interactions that happen after the formation of the MSP.
As discussed in Paper I, none of the MSPs in 47~Tuc is clearly
the product of such an interaction.

Furthermore, in high-$\gamma$ GCs we find many pulsars very far from the cluster
core. An extreme example is NGC~6752 \citep{dpf+02,cpl+06},
a core-collapsed cluster where two of the five known pulsars lie at
more than 14 core radii from the centre. This phenomenon is
common in other core-collapsed GCs and is caused by chaotic
binary interactions, which typically have a strong recoil
that can propel MSPs to the outer reaches of the cluster.
In 47~Tuc, all pulsars but one appear to lie close to the core, their
radial distribution being as expected from mass segregation
of a dynamically relaxed population \citep{hge+05}. Even for the exceptional
system, 47~Tuc~X, it is not clear whether there was a recoil
in the past (Paper I).

A detailed characterization of the pulsar populations of other
GCs will be important for testing this general
picture. The pulsar populations in low-$\gamma$ GCs 
(like M3, M5, M13, M22, M53, M62, NGC 6749 and NGC 6760) should have characteristics
similar to 47~Tuc, thus different from those in the high-$\gamma$ clusters.
This appears to be true
\citep{fhn+05,hrs+07,lrfs11,lfrj12}, but it could be refuted (or further confirmed) by timing
more of the pulsars in those clusters - and finding new ones.

\subsection{An intermediate mass black hole in the centre of 47 Tuc?}

Recently, the possibility of an intermediate mass (2300~$M_{\odot}$) black hole
(IMBH) at the centre of 47 Tuc has been raised considering only the
$\dot{P}_{\rm obs}$ of the pulsars (see \citealt{kbl17} and
associated Corrigendum)
which give us upper limits on the pulsar accelerations ($a_{\ell, \, \rm max}$) via Eq,~\ref{eq:almax}.
In this work, we consider not only the $a_{\ell, \, \rm max}$, but also measurements of
jerk along the line of sight ($\dot{a}_{\ell}$), and actual measurements of the line-of-sight
accelerations in the field of the cluster ($a_{\ell, \, \rm GC}$) for 10 binary pulsars, as
discussed in previous sections. This is important because these accelerations are more
constraining than the $a_{\ell, \, \rm max}$ taken into account in \cite{kbl17}.

The simple analytical cluster model described in 
Section~\ref{sec:jerks} with $d \, = \, 4.69 \, \rm kpc$
can account for all the $a_{\ell,\, \rm GC}$
despite the fact that these are more constraining than the $a_{\ell, \, \rm max}$.
In the cases where these are missing, the model can account for
all the $a_{\ell, \, \rm max}$ (from the $\dot{P}_{\rm obs}$) as well.
Furthermore, the model also
accounts for the jerks observed for all the pulsars that lie (in projection) in the core.
Thus, considering all the available observations, we come to the conclusion that
using our cluster model we find no clear evidence for the existence of an
IMBH at the centre of the cluster: its gravity is not necessary
to explain the observations.

As shown in section~\ref{sec:jerks}, the cluster
distance is crucial for the interpretation of the accelerations.
Once we know the standard deviation of the \emph{HST} proper motions near the centre
of the cluster, then the predicted accelerations are proportional to the
assumed distance. With the smaller distance assumed by \cite{kbl17}, which
we believe to be an under-estimate (see section 1.1), our
cluster model is unable to account for
the observed accelerations of 47~Tuc~S, E and U.

We must, however, emphasize that this discussion is based on our analytical model,
which is not necessarily an accurate description of the actual cluster potential,
particularly if a massive black hole is present.
A probabilistic estimate of the mass of this hypothetical
IMBH will be presented by Abbate et al. (in preparation).

\section{Summary and conclusions}
\label{sec:conclusions}

In this paper, we have reported on the discovery of two millisecond
pulsars in the GC 47~Tucanae (47~Tuc~X and Y) and
presented 20 timing solutions for just as many pulsars in the cluster.
Seventeen of these are updates of previous timing solutions, with
more than 10 years of high-resolution data added, and, in the case of 
47~Tuc~ab, an extended set of ToAs compared to that of \cite{phl+16}.
The remaining three solutions
(for pulsars 47~Tuc~R, Y and Z) are presented here for the first time. 
With the solutions  presented in \cite{rfc+16} (for 47~Tuc~W and X) and in Freire \& Ridolfi(2017)
(for 47~Tuc~aa), we now have a total of 23 timing solutions.

The large timing baseline with uniform coverage from the high-resolution
analogue filterbank at Parkes has
produced a great improvement in the measurement of several key parameters
for these pulsars, in particular those that  are relevant for a study of the
dynamics of the cluster.
The $\dot{P}_{\rm obs}$ have been used in previous studies of
the mass model of 47~Tuc \citep{fck+03}. In this paper we refine
them and present new measurements for pulsars 47~Tuc~R, Y and Z.
Apart from this, we present several new additional pulsar
parameters that are important for a study of the dynamics of the cluster:
the proper motions, the real line-of-sight accelerations
as determined from the orbital period derivatives
and the jerks. All of these will contribute to more detailed analyses
such as that done by \cite{prf+17} for the GC Terzan 5.
These should, in particular, be able to a) link the acceleration
model with the measured proper motions, something we have not
done in this paper, b) provide better probabilistic distance estimates
from the acceleration data
and c) investigate the probability distribution for
the mass of a hypothetical IMBH at the centre of the cluster.

Nevertheless, we can already derive some preliminary conclusions, based on
the stellar proper motion dispersion near the centre of the cluster
and the analytical cluster acceleration model presented in Section~\ref{sec:jerks}:
The measurements of acceleration based on $\dot{P}_{\rm b, \rm obs}$
can be accounted for by this model with a cluster
distance of 4.69 kpc. This coincides with most photometric and
spectroscopic distances published to date, see e.g. \cite{wgk+12}
and references therein; this suggests that the published
$\sigma_0$ and  the kinematic distances are too small.
The likely reasons for this have already been discussed in 
\cite{bhog16} and are summarized in the Introduction. If we instead
use the smaller distances suggested by kinematic studies
then the cluster model is unable to predict the line-of-sight
accelerations of three binary pulsars, 47~Tuc~E, S, and U.

Regarding the jerks, we find that the cluster potential can
also account for most observed jerks, particularly those
in the core. Only 47~Tuc~H, U and J (which lie well outside the core) have
jerks that cannot be explained by any cluster model,
these pulsars are clearly being influenced by nearby objects.

The fact that our analytical model
with $d \, = \, 4.69\, \rm kpc$ can account for all observed
pulsar accelerations and upper limits and all pulsar jerks in the core
means that, with this model, we find no evidence for 
any excess accelerations near the centre of the cluster 
such as should be caused by the presence of an intermediate
mass black hole.

We have also described the behaviour of the four black widow systems
with known timing solutions, 47~Tuc~I, J, O and R. Although two of the
systems (47~Tuc~J and O) exhibit strong variability in their
orbital periods, as observed in the long-term timing of
other such systems \citep{svf+16},
two others (47~Tuc~I and R) appear to be stable, with
orbital period derivatives that are very similar to
those expected from their acceleration in the field of the cluster.
This hints at a bi-modal behaviour of the black widow systems.
If confirmed by the study of other systems, this is important
for a variety of applications: some black widows might be 
suitable for use in pulsar timing arrays.

We can now derive an improved value for the total mass of 47~Tuc~H.
However, it is not yet possible to measure the masses of the
individual components of that system. Furthermore, we have detected the
rate of advance of periastron for three more systems (47~Tuc~E, S and U),
but these are not yet precise enough for astrophysically
interesting constraints on the total mass.

% \subsection{Prospects for future timing}

% Continued timing of the known pulsars will refine and increase the number of proper motions, possibly enlarging the current velocity envelope and shifting its centre, it will greatly improve the measurements of $\dot{P}_b$ and refine the measurements of $\dot{P}_{\rm int}$, it will allow a better characterization of the "black widow" systems (and confirm or refute the idea that some are stable), refine and increase the number of mass measurements - with a few potential NS mass measurements - and finally determine whether 47~Tuc~H and U are in triple systems. The long-term timing of the pulsars in 47~Tuc will likely remain scientifically rewarding.

\section*{Acknowledgements}
PCCF and AR gratefully acknowledge financial support by the European Research Council
for the European Research Council Starting grant
BEACON under contract no. 279702, and continuing support from the Max Planck Society.
AR is member of the International Max Planck research school for Astronomy and
Astrophysics at the Universities of Bonn and Cologne and acknowledges partial support through the
Bonn-Cologne Graduate School of Physics and Astronomy. PT gratefully acknowledges
financial support from the European Research Council for the
SynergyGrant BlackHoleCam (ERC-2013-SyG, Grant Agreement no. 610058).
COH acknowledges funding from an Natural Sciences and Engineering Research Council of
Canada Discovery Grant, a
Discovery Accelerator Supplement, and an Alexander von Humboldt Fellowship.
The Parkes radio telescope is part of the Australia Telescope, which is funded by the
Commonwealth of Australia for operation as a National Facility managed by the Commonwealth
Scientific and Industrial Research Organisation.
DRL was supported by National Science Foundation RII Track I award number 1458952.
We especially thank Ralph Eatough for observations, significant assistance with the
data handling and for stimulating discussions on pulsar searches,
Norbert Wex for important comments and help with the whole project
and the referee Scott Ransom for carefully reviewing 
this work and providing suggestions that improved and shortened it.
This research has made use of NASA's Astrophysics Data System.

%%%%%%%%%%%%%%%%%%%%%%%%%%%%%%%%%%%%%%%%%%%%%%%%%%



%%%%%%%%%%%%%%%%%%%%%%%%%%%%%%%%%%%%%%%%%%%%%%%%%%

%%%%%%%%%%%%%%%%% APPENDICES %%%%%%%%%%%%%%%%%%%%%

% \appendix

% \section{Some extra material}

% If you want to present additional material which would interrupt the flow of the main paper,
% it can be placed in an Appendix which appears after the list of references.

%%%%%%%%%%%%%%%%%%%%%%%%%%%%%%%%%%%%%%%%%%%%%%%%%%

% Don't change these lines
\bsp	% typesetting comment
\label{lastpage}
\end{document}